\documentclass[floats,floatfix,showpacs,amssymb,amsmath,prd,superscriptaddress,nofootinbib, aps,twocolumn]{revtex4-2}

\usepackage[utf8]{inputenc}
\usepackage[T1]{fontenc}
\usepackage{hyperref}
\usepackage{bm}
\usepackage{dcolumn}
\usepackage{aas_macros}
\usepackage{amsmath, amssymb}

\usepackage{graphicx}
\usepackage{float}
\usepackage[normalem]{ulem}
\usepackage{soul}
\setstcolor{red}
\hypersetup{colorlinks,bookmarksnumbered,
citecolor=cyan, urlcolor=magenta}

\providecommand{\adsurl}[1]{\href{#1}{ADS}}
\usepackage{natbib}
\usepackage[usenames, dvipsnames]{xcolor}

\usepackage{array}
\usepackage{rotating}
\newcolumntype{C}[1]{>{\centering\arraybackslash}m{#1}}

\begin{document}
\preprint{APS/123-QED}

\title{Tolerance to Astrophysical Model Uncertainty in Dark Siren Hubble Measurement with Third-generation Gravitational-wave Detectors}
\author{Yijun Wang}
\email{yijunw@caltech.edu}
\affiliation{California Institute of Technology, Pasadena, CA 91125, USA}
\author{Yanbei Chen}
\affiliation{California Institute of Technology, Pasadena, CA 91125, USA}

\date{\today}

\begin{abstract}

Gravitational-wave (GW) events can serve as standard sirens for cosmology, as the luminosity distance to source can be directly measured from the waveform amplitude. Specifically, the ``dark'' siren method involves inferring cosmological parameters, e.g. the Hubble constant, by comparing the luminosity distance distribution and that of the redshift, typically obtained through a combination of galaxy survey catalog and theoretical models. Especially with the prospect of third-generation GW detectors, the statistical uncertainty of the Hubble measurement can be suppressed to a percent level. However, incorrect assumption in galaxy population models can lead to systematic bias, which becomes increasingly relevant as third-generation GW detectors can detect large-redshift sources beyond the reach of currently available galaxy catalogs. In this work, we adopt a Fisher information formalism and study the maximum model error tolerance given specific total error budget. We find that, to achieve a total error budget of $1\%$ in the Hubble constant, the galaxy mass function redshift evolution should be known to $\mathcal{O}(1\%)$. We find that galaxy redshift uncertainty, survey magnitude limit and GW angular localization error are important factors. We find galaxy clustering also improves model error tolerance, thus our results provide a conservative benchmark to future analysis using real galaxy catalogs. We also investigate the effective bright siren scenario and highlight that the dark siren selection strategy should be catered to measurement uncertainty and the target total error budget. This work thus highlights the challenge in the dark siren method and quantifies requirements on both the galaxy catalog and GW measurement to contribute to constraining cosmology. 

\end{abstract}
\maketitle

\section{Introduction}
\label{chap:intro}
A fundamental topic in modern cosmology is the understanding of the growth history of the universe. In particular, the $\Lambda$ Cold Dark Matter ($\Lambda$CDM) model has proven to be very successful with current cosmological and astrophysical observations \cite{Dodelson}. The values of the cosmological parameters in this model are constrained with modern observations to increasing level of precision, such as that of the Hubble constant. However, inconsistency of the Hubble constant values estimated with different methods appears, which is famously known as the Hubble tension. 

On one hand, the Hubble constant can be constrained from early-universe measurement on the cosmic microwave background (CMB) power spectrum. The Planck survey on CMB yields a Hubble constant of $H_0=67.4\pm 0.5$ km/s/Mpc \cite{Planck18}. An independent late-universe measurement relies on ``standard candles'' \cite{Riess1996,deJaeger,Dhawan2018} with known luminosity such as Cepheids in Type Ia supernovae hosts. For example, Milky Way Cepheids in the early data release 3 (EDR3) of \textit{Gaia} measure a Hubble constant of $73.2\pm 1.3$ km/s/Mpc \cite{Riess2021}. Evidently, the two methods yield estimates that are $4.2\sigma$ apart, posing a strong challenge if we require consistency within the same model \cite{Verde,Kamionkowski}. 

As gravitational-wave (GW) astronomy advances with successful event detections \cite{GW150914,GW170104,GW170817,GW190521}, the possibility to use GW events as standard sirens for studying the Hubble constant draws attention \cite{Schutz,Chen,gwtc2_hubble}. Specifically, the GW source luminosity distance can be inferred from the waveform; if its redshift information becomes available, the Hubble constant and other cosmological parameters can be constrained.

In the simplest case, the GW event is accompanied by an electromagnetic counterpart, and its redshift is determined with an EM telescope. In this case, the GW standard siren is also called a ``bright'' siren. The sole example to date is GW170817 \cite{GW170817}, a binary neutron star merger. Its gamma-ray burst signal was registered independently by the Fermi Gamma-ray Burst Monitor, and the Anti-Coincidence Shield at the International Gamma-Ray Astrophysics Laboratory (INTEGRAL) spectrometer\cite{gw170817_grb}. This event alone constrained the Hubble constant to $70^{+12}_{-8}$ km/s/Mpc \cite{gw170817_hubble}. The bright sirens do not represent the full potential of standard sirens, as they make up only a small fraction of the total GW events. To start with, GW source population synthesis studies show that binary neutron star (BNS) events are fewer than binary black hole (BBH) events by a factor of $\mathcal{O}(10)$ \cite{Biesiada,BNS_aLIGO}. Among the detectable BNS events, only $0.1\%$ is estimated to have observable EM counterpart \cite{Califano}. 

While individual GW events without EM counterpart cannot constrain cosmological parameters due to missing redshift, the total detected population could be informative in statistical inference of cosmological parameters \cite{Schutz}. This method is also called the GW dark siren. In short, the detected catalog can be compared to an ``expected'' distribution as a function of the cosmology and the astrophysics \cite{Hang}. Depending on the main focus parameter space, we can further define a sub-category called the ``spectral siren'', where the focus is primarily on comparing source mass distribution \cite{Chernoff,Farr_spectral,Ezquiaga_spectral_siren,Farah}. In this work, we focus on the redshift distribution, which is derived from galaxy distribution with the assumption that GW events are biased tracers of galaxies \cite{Gray2020,Gray2022,Finke}. 

Such method has been applied to GW catalogs; in the Third LIGO-Virgo-KAGRA Gravitational-Wave Transient Catalog (GWTC-3), 47 events are analyzed against the galaxy catalog GLADE+ \cite{glade}, and together with GW170817 yield $H_0=68^{+8}_{-6}$ km/s/Mpc \cite{Abbott_gwtc3}. While the current detector sensitivity limits the application of this method (see, e.g., Fig 7 in Ref. \cite{Gray2022}), it is predicted that current detector upgrades and third-generation detectors can produce GW catalog that constrains the Hubble constant towards the percent level \cite{Chen,Zhu_Tianqin,Shafieloo,Alfradique}. 

In the dark siren method, the expected distribution is typically a mixture model between a galaxy catalog and theoretical population models \cite{Gray2020,Gray2022,Finke,Hang}. This treatment is necessary, since galaxy catalogs can be incomplete due to, e.g., telescope magnitude limits. Since the relation between GW events and host galaxy characteristics is still an active area of research \cite{Fishbach2018}, the detected events can well come from galaxies that are missing from catalog. In this case, the redshift information is known only to the degree of the smooth overall galaxy mass function. It is evident that, with an incorrect theoretical model, the inferred cosmology is susceptible to bias. In Ref. \cite{gwtc2_hubble}, the effect of varying the event rate redshift evolution is explored; however, for this catalog, the Hubble constant inference is dominated by the bright siren GW170817, which is not affected by astrophysical model assumption (see Figure 7 in Ref. \cite{gwtc2_hubble}). In Ref. \cite{Hang}, bias incurred by elemental astrophysical model substructures (modeled as gaussian peaks) is illustrated. 

In this work, we seek to quantify the susceptible bias using a Fisher information framework. Compared with Monte Carlo simulations (e.g., \cite{Gray2020,Gray2022}), this method is more cost-efficient and produces quick estimates of the bias. Compared with Ref. \cite{Hang}, this paper considers realistic mixture models with simulated galaxy catalogs, thus emphasizing the interplay between galaxy survey precision and GW catalog selection. 

This paper is organized as follows; in Section \ref{chap:theory}, we show the Fisher information framework. In Section \ref{chap:sim}, we highlight how galaxy catalog and GW measurement uncertainties are factored into our simulation and introduce model variations. In Section \ref{chap:analysis} and \ref{sec:beta_analysis}, we examine the simulation cases and explain results on galaxy model error tolerance. In Section \ref{chap:conclusion}, we summarize our findings and outline directions for further studies. 

\section{Fisher for Dark Siren} 
\label{chap:theory}
In this section, we specify the Fisher formalism for calculating statistical and systematic error on the Hubble constant. As discussed in Section \ref{chap:intro}, the statistical dark siren method amounts to comparing the detected GW population to an expected rates model, $r(\hat{D},\hat{\Theta})$, where $\hat{\Theta}$ is a vector of parameters including, e.g., the sky location, $\hat{\Omega}$, and redshifted mass parameter of the binary, $(\hat{m}_1,\hat{m}_2)$. Throughout this work, we use $\hat{\cdot}$ for detector-frame GW quantities, $\tilde{\cdot}$ for galaxy catalog quantities, and no additional superscripts for the underlying true distribution. We also focus on estimating the Hubble constant alone and assume a standard $\Lambda$CDM cosmology with $H_0=70$ km/s/Mpc and $\Omega_m=0.3$.

We discuss the mass contribution by the end of this section, and we focus on the luminosity distance and angular sky location, $\hat{D},\hat{\Omega}$, for now. The expected rate of GW events is given by 
\begin{equation}
    r_\mathrm{det}(\hat{D},\hat{\Omega}) = f_\mathrm{ gw}(\hat{D},\hat{\Omega})r(\hat{D},\hat{\Omega})\;,
\label{eqn:gw_compfrac}
\end{equation}
where $f_\mathrm{ gw}$ is the completeness fraction of GW detection, i.e., for sources at $\hat{D}$, $\hat{\Omega}$, what fraction of sources exist in the catalog.

The full rate, $r(\hat{D},\hat{\Omega})$, is a convolution of the underlying GW event population, $r(D,\Omega)$, with the detector sensitivity kernel. We model the luminosity distance direction and angular direction sensitivity kernel as a gaussian distribution with $(\hat{D},\hat{\sigma}_D)$ and a three dimensional von Mises-Fisher distribution with parameter $\hat{\kappa}$. They are assumed to be independent, and their joint distribution is their product.

The underlying full rate can be written as a mixture model between an empirical part from galaxy catalog and a supplementary theoretical component. The proportion depends on the completeness fraction of the catalog, $f_\mathrm{g}$. For magnitude-limited surveys, the completeness fraction is computed from distribution of galaxy absolute magnitude and the luminosity distance. 

We assume that the galaxy catalog is in the form of $(z_i, \sigma_i, \Omega_i)$ representing the measured redshift, redshift uncertainty and angular position of galaxy $i$. The uncertainty kernel shares a similar form as the GW kernel, except that the gaussian distribution is given in redshift, and the parameter $\kappa_i$ is much larger than $\hat{\kappa}$. The rate of GW events in galaxy $i$ is given by $r_i^g$. The true galaxy mass function is written as $r_\mathrm{ true}(z)$. Performing analytical integrals, we obtain

\begin{widetext}
\begin{equation}
\begin{split}
    r(\hat{D},\hat{\Omega}) &= \int r(D,\Omega)K(D,\hat{D},\Omega,\hat{\Omega})dDd\Omega\\
    &=\int [r_\mathrm{ cat}(D,\Omega)+(1-f_g(D))r_\mathrm{ true}(D,\Omega)]K(D,\hat{D},\Omega,\hat{\Omega})dDd\Omega\\
    &\equiv r_\mathrm{ cat}(\hat{D},\hat{\Omega})+r_\mathrm{ theo}(\hat{D})\;,
\end{split}
\label{eqn:rfulldef}
\end{equation}

\begin{equation}
    r_\mathrm{ cat}(\hat{D},\hat{\Omega}) = \sum_i \frac{r_i^g}{\sigma_i\hat{\sigma}_D} \frac{\hat{\kappa}}{(2\pi)^2}e^{\hat{\kappa}(\Omega_i\cdot\hat{\Omega}-1)}\int\exp\left[-\frac{(z-z_i)^2}{2\sigma_i^2}\right]\exp\left[-\frac{(D(z)-\hat{D})^2}{2\hat{\sigma}_D^2}\right]dz\;,
\label{eqn:rcat}
\end{equation}

\begin{equation}
    r_\mathrm{ theo}(\hat{D}) = \frac{1}{\sqrt{2\pi}\hat{\sigma}_D}\int [1-f_g(D(z))]r_\mathrm{ true}(z)\exp\left[-\frac{(D(z)-\hat{D})^2}{2\hat{\sigma}_D^2}\right]dz\;,
\label{eqn:rtheo}
\end{equation}
\end{widetext}

where we separate into the catalog-piece and theoretical piece of the rate in the last line of Eqn. \eqref{eqn:rfulldef}.

Treating the detected GW events as Poisson samples from this rates model, the Fisher information can be written as \cite{Hang}
\begin{equation}
    \mathcal{I}_{H_0} = \int r_\mathrm{ det}(\hat{D},\hat{\Omega})\left(\frac{d \log r_\mathrm{ det}}{d H_0}\right)^2 d\hat{D}d\hat{\Omega}  \;,
\label{eqn:fisher_info}
\end{equation}
and the statistical uncertainty is given by 
\begin{equation}
    \delta H_{0,\mathrm{s}} = 1/\sqrt{\mathcal{I}_{H_0}}\;.
\label{eqn:dHs}
\end{equation}

Consider any rate model $q(D)$ derived from a fixed redshift rate $q(z)$ with $dz/dD$. The Hubble constant derivative is given as 
\begin{equation}
    \begin{split}
    \frac{d\log q(D)}{d H_0}|_D = -\frac{1}{H_0}\left(\frac{d\log q(D)}{d\log D} + 1\right)
    \end{split}\;,
\end{equation}
which can be shown by applying the chain rule. This result can be directly applied to $r_\mathrm{ cat}(D,\Omega)$ and the integrand of $r_\mathrm{ theo}$. Since the completeness fractions are functions of luminosity distance in the detector frame, they do not change under variation of $H_0$. Accordingly, 
\begin{equation}
    \frac{d \log r_\mathrm{ det}(\hat{D},\hat{\Omega})}{d H_0} = \frac{d \log r(\hat{D},\hat{\Omega})}{d H_0}\;.
\end{equation}
The Hubble constant derivative of the detected GW rate is then given by 

\begin{equation}
    \frac{d\log r_\mathrm{ det}(\hat{D},\hat{\Omega})}{dH_0} = \frac{1}{r(\hat{D},\hat{\Omega})}(I_1+I_2)
\label{eqn:drv}
\end{equation}
\begin{equation}
    \begin{split}
        I_1 & = -\frac{1}{H_0}r_\mathrm{ cat}(\hat{D},\hat{\Omega}) -\frac{1}{H_0}J_1\\
        I_2 & = -\frac{1}{H_0}r_\mathrm{ theo}(\hat{D}) -\frac{1}{H_0}J_2
    \end{split}
\end{equation}
\begin{widetext}
    \begin{equation}
    J_1 = \sum_i \frac{r_i^g}{\sigma_i\hat{\sigma}_D} \frac{\hat{\kappa}}{(2\pi)^2}e^{\hat{\kappa}(\Omega_i\cdot\hat{\Omega}-1)}\int\exp\left[-\frac{(z-z_i)^2}{2\sigma_i^2}\right]\frac{z_i-z}{\sigma_i^2}\exp\left[-\frac{(D(z)-\hat{D})^2}{2\hat{\sigma}_D^2}\right]D(z)\frac{dz}{dD}dz
\end{equation}
\begin{equation}
    J_2 = \frac{1}{\sqrt{2\pi}\hat{\sigma}_D}\int [1-f_g(D(z))]\frac{dr_\mathrm{ true}(z)}{dz}\exp\left[-\frac{(D(z)-\hat{D})^2}{2\hat{\sigma}_D^2}\right]D(z)\frac{dz}{dD}dz\;.
\end{equation}
\end{widetext}

The derivative of $r_\mathrm{true}$ can be analytical if the redshift evolution is given by some parametric function of redshift; in this work we use finite differencing to approximate its value. 

This Fisher formalism can also be adapted to estimate bias from the true Hubble constant value, if the assumed models deviate from the true model, $\tilde{r}$. For instance, both the theoretical galaxy mass function, $r_\mathrm{ true}(z)$, and the GW event bias, $r_i^g$, are active areas of research. Semi-analytic models and hydrodynamic simulations on galaxy redshift evolution attempt to include various physical processes such as gas cooling, star formation, feedback, etc., but the model details vary \cite{Torrey}. GW event bias\footnote{In this work, the most common usage of ``bias'' refers to systematic errors in contrast to statistical ones. In a few places, such as the present location, it is inherited from cosmology studies and denotes that GW sources are tracers of galaxies.} can also be redshift dependent to reflect increase in merger efficiency in low metallicity galaxy environment \cite{Fishbach2018}. In the most general form, the deviation can be written as 
\begin{equation}
    \tilde{r}(\hat{D},\hat{\Omega}) = r(\hat{D},\hat{\Omega}) + \delta r(\hat{D},\hat{\Omega})\;.
\end{equation}
The resulting bias on the Hubble constant can be shown to be \cite{Hang}
\begin{equation}
    \delta H_{0,\mathrm{b}} = \frac{1}{\mathcal{I}_{H_0}}\int \frac{d \log r(\hat{D},\hat{\Omega})}{dH_0}\delta r d\hat{D}d\hat{\Omega}\;.
\label{eqn:bias}
\end{equation}
While the statistical uncertainty scales as $1/\sqrt{T}$, where $T$ is the total observation time, the error term $\delta r$ is also proportional to $T$ and the bias stays constant. Especially in the context of third-generation detectors where we expect large numbers of high precision GW events, the role of the bias will be increasingly important in determining the level of total error budget. 

Since GW measurement probes redshifted mass rather than the intrinsic mass, the event rate dependence on redshifted mass is also subject to changes in the Hubble constant. Specifically, the full rate model with explicit mass dependence is given by 
\begin{widetext}
    \begin{equation}
\begin{split}
    \frac{\partial N}{\partial \hat{M}\partial \hat{D}} &= \int \frac{\partial N}{\partial M\partial D}P(M|\hat{M})P(D|\hat{D}) dDdM\\
    &=  \int \frac{\partial N}{\partial m\partial D}\frac{1}{1+z(D)}P(M|\hat{M})P(D|\hat{D}) dD(1+z(D))dm\\
    & = \int \frac{dN}{dD}p(m)P(M|\hat{M})P(D|\hat{D}) dDdm\\
    & = \int \left(\int \frac{dN}{dD}(m(1+z(D)), D)P(M=m(1+z(D))|\hat{M})P(D|\hat{D}) dD\right)p(m)dm\;.
\end{split}
\label{eqn:dNdMdD}
\end{equation}
\end{widetext}

The Hubble constant derivative of this piece with mass information is given as 
\begin{widetext}
    \begin{equation}
    \begin{split}
        \frac{d}{dH_0}\left(\frac{\partial N}{\partial D \partial M}\right) & = \frac{d}{dH_0}\left(\frac{\partial N}{\partial D \partial m}\frac{1}{1+z(D)}\right)\\
        & = \frac{d}{dH_0}\left(\frac{d N}{dD}p(m)\frac{1}{1+z(D)}\right)\\
        & = \left(\frac{d}{dH_0}\frac{dN}{dD}\right)p(m)\frac{1}{1+z(D)} + \frac{dN}{dD}p(m)\frac{1}{(1+z(D))^2}\frac{D}{H_0}\frac{dz}{dD}\;,
    \end{split}
\end{equation}
\end{widetext}

where the rate and luminosity distance direction derivatives are given in Eqn. \eqref{eqn:rfulldef} and Eqn. \eqref{eqn:drv}.

\section{Simulation}
\label{chap:sim}
In this section, we explain various components in our simulation and enumerate parameters of the model cases. Firstly, we describe the simulation volume. With the desirable GW angular uncertainty in mind, we simulate a cone with radius $\theta_\mathrm{max}=4$ deg using the \texttt{Healpy} package in Python \cite{healpix,healpy}. We set the resolution parameter $\texttt{NSIDE}=2^9$, which gives 3784 equal-areal pixels within the simulated region. The redshift ranges from $0.01$ to $2$. The lower bound ($\sim 43$ Mpc) is sufficiently small to include the bulk expected GW population. The upper bound $z=2$ is picked in light of realistic galaxy catalog depth; even though third-generation GW detectors may detect significant events at even higher redshift, the host galaxy information can be so incomplete that the dark siren becomes subject to above-tolerance bias. We assume that all-sky data is available for both galaxy catalog and GW catalog, and we further assume that all sky patches are statistically similar, including galaxy and GW event rates, distribution and measurement quality. Since the Fisher information scales linearly with angular coverage, we add a multiplicative factor equal to the fraction of simulated area compared to the whole sky. We note that the potential bias, $\delta H_{0,b}$, is independent from sky coverage. This suggests that the all-sky case is the most optimistic in terms of reducing statistical error and consequently the total error budget. 

We create two broad groups of models that treat GW measurement differently. In the first group, which we refer to as the fiducial models, GW uncertainties are modeled using simple scaling functions with redshift. In the second group, GW uncertainties are computed from Fisher matrix with realistic third-generation detector network sensitivity. We refer to the latter as realistic models. In the following sections, we go into details about the setup for each.

\subsection{Galaxy Simulation}

We now discuss simulation parameters related to the galaxy catalog. There are three relevant aspects of galaxy catalog, the measurement quality, depth and overall rates, and we discuss each in turn.

\subsubsection{Galaxy Redshift Measurement Precision}

In our simulation, we consider two levels of redshift measurement precision with the form $\sigma_i(z) = \sigma_{i,0}(1+z)$ (recall that the index $i$ refer to galaxy index in the catalog). In most fiducial cases and all cases with Fisher uncertainty, we set $\sigma_{i,0}=0.003$, corresponding to typical value for spectroscopic surveys. For photometric survey redshift measurement, the target redshift error is $\sigma_{i,0}=0.01$. We explore the effect of larger redshift uncertainty with the model labeled \textit{\texttt{photo}}.

\begin{figure}[hbt!]
    \centering
    \includegraphics[width=\columnwidth]{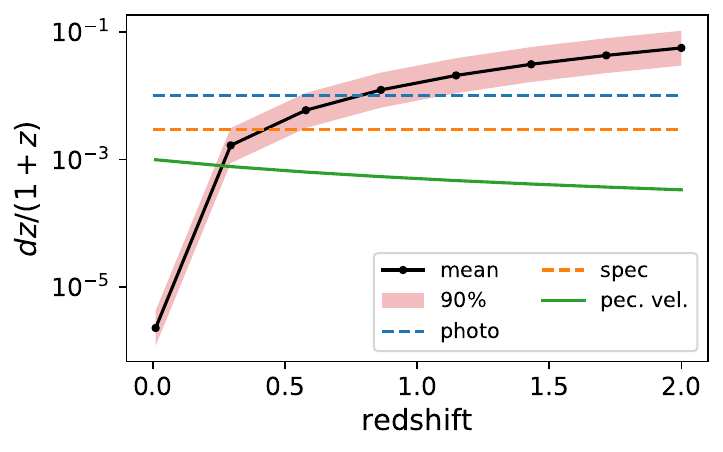}
    \caption{Redshift error budget comparison between galaxy measurement, GW measurement and peculiar velocity. The luminosity distance error is computed from a binary with intrinsic mass $(m_1,m_2)=(8.28~M_\odot,7.42~M_\odot)$, and the light red band marks the 90\% percentile across sky locations. }
    \label{fig_pecvel}
\end{figure}
%For close-by galaxies, the peculiar velocity can be a significant contribution to the redshift error budget. In our case, it is necessary to include more sources with deeper redshift to suppress statistical uncertainties. 

Aside from galaxy survey measurement error, the galaxy peculiar velocity also introduces error on the inferred redshift. In Ref. \cite{Mukherjee_pecvel,Howlett_pecvel,Nicolaou,Nimonkar}, the authors studied the effect of considering galaxy-mass-dependent peculiar velocity in GW bright sirens, which are attainable at small redshift. In Ref. \cite{Zhu_Tianqin}, the peculiar velocity is included as a constant in galaxy redshift uncertainty. In Figure \ref{fig_pecvel}, we compare the error from galaxy measurement, GW measurement and peculiar velocity. We show both spectroscopic and photometric redshift uncertainty value. For peculiar velocity, we adopt a typical value of 500 km/s \cite{Zhu_Tianqin}. We use the reference GW source with intrinsic mass $(m_1,m_2)=(8.28~M_\odot,7.42~M_\odot)$ and compute the volume-weighted luminosity distance uncertainty (see below for details). The light red block represents the 90 percentile over sky locations. We make a few observations. Firstly, at all redshifts, the peculiar velocity gives smaller error than the nominal spectroscopic and photometric uncertainties, especially with increasing redshift. This suggests that, given our galaxy survey assumptions, it is reasonable to ignore peculiar velocity and focus on the main source of error. This assumption, however, is expected to break down for very small redshift galaxies, since their magnitude allows for better quality measurement than the typical values quoted here. If a more realistic or a real galaxy catalog is used, it is necessary to re-evaluate and incorporate peculiar velocity. In the present study, we will neglect the contribution from peculiar velocity. We also observe that, the redshift budget from GW contribution is in general smaller than galaxy contribution up until $z\sim 1$. For the assumed galaxy magnitude limit, the completeness fraction is typically already quite small at these redshifts (e.g. Figure \ref{fig_fisher_all_zslice} and Figure \ref{fig_fid_all_zslice}). This comparison suggests that in the radial direction, the dominating source of uncertainty comes from galaxy catalogs rather than from GWs with third-generation sensitivity. 

\subsubsection{Galaxy Survey Completeness Fraction}

The depth of a galaxy survey is related to many factors, such as the target galaxy magnitude or presence of special emission lines. For example, one of the targets of the Dark Energy Spectroscopic Instrument (DESI) \cite{desi_white} is Luminous Red Galaxies (LRG). Such galaxies exhibit prominent 4000\AA ~spectral break, and they will be the primary observation targets for DESI in redshift range $0.4<z< 1.0$ \cite{Zhou_lrg}. In the redshift range of $0.6<z<1.6$, another group of important targets are the Emission Line Galaxies (ELG), and the spectral signature is the [O II] doublet \cite{Raichoor_elg}. Since GW event distribution among different galaxy types is still an ongoing area of research \cite{cao_hostgalaxy}, we make the simplifying assumption that the GW bias, i.e., its relative density per galaxy, is uniform. Under this premise, although galaxy surveys that target specific emission lines can reach large redshift, the catalog sample excludes galaxies without these signatures, and can be consequently quite incomplete for our purpose. 

Instead, we consider a magnitude-limited galaxy catalog such as the Bright Galaxy Survey (BGS) \cite{desi_bgs} of DESI. The DESI BGS is especially promising due to its large sky coverage and its spectroscopic redshift precision. We note that DESI BGS is expected to have a $\sim$14,000 deg$^2$ footprint \cite{desi_bgs}, which is not fully consistent with our all-sky catalog assumption. We expect that other surveys focusing on different sky patches can fill each other's gaps and provide a joint catalog (e.g. see GLADE \cite{glade}). For example, the Wide Survey of Euclid \cite{euclid_def} is expected to cover $\sim$15,000 deg$^2$ that complements the footprint of DESI. This survey is also expected to offer photometric redshift error with $\sigma_i/(1+z)<0.05$ and spectroscopic measurement with $\sigma_i/(1+z)<0.01$ \cite{euclid_def}. Consequently, we neglect sky coverage details and quote only the instrumentation magnitude limit of DESI BGS in our simulation. 

The magnitude-limited completeness fraction depends on the absolute magnitude distribution, which can be given as empirical fits from local surveys in the Schechter form, \cite{Abbott2021,Gray2020,Gray2022}
\begin{equation}
    p(M)\propto 10^{-0.4(\alpha_M+1)(M-M^*)}\exp\left[-10^{-0.4(M-M^*)}\right]\;.
\end{equation}
For the shape parameters, we adopt the empirical fits in Ref. \cite{Blanton2003} derived from local galaxy catalog from SDSS near $z=0.1$. For $r$-band magnitude, the best-fit parameters are given by $M^* = -20.44+5\log_{10}h,\alpha_M=-1.05$, where $h$ is the Hubble parameter. The absolute magnitude limits are given by $(-24.26,-16.11)$. The absolute magnitude is converted to the apparent magnitude $m$ with $m = M+5\log_{10}(D/10~\mathrm{ pc})$. Integrating up to the magnitude limit of the galaxy survey, we obtain the corresponding completeness fraction. 

In realistic surveys, the observation completeness can be further dependent on more instrumentation and scheduling details. During DESI BGS, the Bright objects ($r$-band magnitude $r<19.5$) are top-priority objects and are observed with high completeness; after four telescope passes, $>80\%$ of BGS Bright targets are assigned a fiber for spectroscopic measurement. The majority of the Faint objects ($r$-band magnitude $r<20.175$) has a lower priority, and they only achieve a fiber assignment efficiency of $60\%$ \cite{desi_bgs} with four passes. The fiber assignment efficiency counts towards the galaxy completeness fraction in addition to the magnitude limit. However, this scaling is dependent on specific observation priorities and scheduling strategy, and we neglect it in this present study for the sake of generality.

In our models, we mostly adopt an apparent magnitude limit of $m_{g}=20.175$ as an optimistic baseline, corresponding to the Faint targets in DESI BGS \cite{desi_bgs}. In the fiducial model named \textit{\texttt{bright}}, we set $m_g=19.5$ to simulate a shallower catalog (mimicking the DESI Brights target). We also set $m_g=22$ for the fiducial model \textit{\texttt{deepz}} to assess the performance gain with a deeper survey.

\subsubsection{Galaxy Number Density}
We compute the galaxy number density using galaxy mass function fitted from observation. For the galaxy number distribution, $\Psi(\sigma_v,z)$, where $\sigma_v$ is the surface velocity dispersion of the galaxy, we first adopt a modified Schechter function fitted from the Sloan Digital Sky Survey (SDSS) catalog \cite{Choi},
\begin{equation}
    \Psi(\sigma_v,0) = \phi_*\left(\frac{\sigma_v}{\sigma_*}\right)^\alpha \mathrm{{exp}}\left[-\left(\frac{\sigma_v}{\sigma_*}\right)^\beta\right]\frac{\beta}{\sigma_v\Gamma\left(\alpha/\beta\right)}\;,
\end{equation}
where $\phi_* = 8.0\times10^{-3}h^3~\mathrm{Mpc}^{-3}$, $\sigma_*=161~\mathrm{km/s}$, $\alpha=2.32$ and $\beta=2.67$. To account for the redshift dependence, we apply a multiplicative scaling factor,
\begin{equation}
    \Psi(\sigma_v,z) = \Psi(\sigma_v,0)\frac{\Psi_\mathrm{{hyd}}(\sigma_v,z)}{\Psi_\mathrm{{hyd}}(\sigma_v,0)}\;,
\end{equation}
where $\Psi_\mathrm{{hyd}}(\sigma_v,z)$ is the fitted function from hydrodynamical simulations in Ref. \cite{Torrey}. The integrated surface velocity dispersion ranges from $\sigma_v=100$ to 300 km/s to include the peak of this distribution. We note that this Schechter function form is fitted from a selection of early-type galaxy samples. Since the distribution of GW sources are not limited to massive early-type galaxies, this rate might be an underestimate. Moreover, with improved galaxy survey sensitivity, the small $\sigma_v$ end could improve from more measurement on faint galaxies. As a sanity check, the resulting galaxy number density at the smallest redshift in our simulation is $0.002$ Mpc$^{-3}$ (i.e. average comoving distance separation is roughly 7.9 Mpc.). To gauge the effect of changes in galaxy number density, we set a fiducial model with reduced galaxy density, labeled \textit{\texttt{sparse}}.

\subsubsection{Catalog generation}
Combining the intrinsic galaxy mass function and the observational completeness, we obtain the expected rates of galaxies in catalog. In this work, the galaxy catalogs are mainly generated via Poisson sampling with even angular distribution over our simulated sky patch. We note that we do not further distinguish the generated galaxy samples, e.g., their stellar mass etc. It is possible to assign galaxies different hosting rates, $r_i^g$, which alters the distribution shape. Since this variation is a straightforward adaptation to the current model, we keep our model general and exclude highly-model-dependent modifications. We also note that the Fisher information from the catalog piece can vary based on the exact catalog realization. However, due to the large number of expected galaxies, the total Fisher information is very stable; for the \textit{\texttt{optimistic}} model parameters (see Table \ref{tables_fid_catalog_params}), we generate 50 catalog realizations and find that the Fisher information standard deviation is $0.053\%$ of the average. For computational efficiency concerns, in the final catalogs reported here, we use only a single galaxy realization for each pair of galaxy magnitude limit and overall galaxy density, $(m_g, f_\mathrm{g,red})$. We list the number of simulated galaxies for each case in Table. \ref{tables_ngal}.

We further explore the effect of galaxy clustering on the realistic cases using the following method. We first query the simulated galaxy catalog from the MICE grand challenge lightcone simulation, \texttt{MICECATv2.0},  \cite{MICE_dm_cluster,MICE_galaxy} through the platform CosmoHub \cite{cosmohub_1, cosmohub_2}. The queried region is a cone centered on (ra,dec) = $(45^\circ, 45^\circ)$ with a radius of 4 deg. Given our reference galaxy survey observational depth, we include galaxies out to $z=0.8$. Specifically, we use the data columns \texttt{ra\_gal}, \texttt{dec\_gal} and \texttt{z\_cgal} and exclude the effect of galaxy peculiar velocity and weak lensing to keep consistent within this paper. We note that the galaxy mass function in the simulation is not identical to the Schechter model used in the majority of this work, and generally contains more galaxies than our assumption. To isolate the effect of clustering from the number density redshift distribution, we resample the MICE catalog; the selection probability is given by the ratio between the queried number and the Schechter model, multiplied by the galaxy observational completeness. The MICE catalog is assumed to be complete within our queried region \cite{MICE_galaxy}.  

\subsection{Gravitational-wave Simulation}

Similar to galaxy simulation, inputs from GWs include measurement quality, completeness fraction and overall event rate. As mentioned, our test models are divided into two broad categories; for realistic models, the test GW event is drawn from an observationally constrained BBH mass function, and the measurement uncertainties are computed from Fisher information with a third-generation detector network. In the following sections, we go through each factor. 

\subsubsection{GW measurement quality}

In this section, we discuss measurement error in luminosity distance and angular error. In all cases, the GW redshift kernel is modeled as a gaussian with $(\hat{D},\hat{\sigma}_D)$. The angular kernel is modeled with the von Mises-Fisher distribution with $p=3$,
\begin{equation}
    p(\Omega,\hat{\Omega}) = C(\kappa) e^{\kappa\cos\Omega\cdot\hat{\Omega}},~~C(\kappa)=\frac{\kappa}{4\pi\sinh\kappa}\;,
\end{equation}
where $\kappa$ controls the spread of the distribution. The value of $\kappa$ is set such that $p(1\sigma)^2$ of the probability is contained within a circular cap with area equal to $\sin\theta \Delta\theta\Delta\phi$, where $\Delta\theta,\Delta\phi$ are the polar angle and azimuthal angle uncertainty of the GW event. 

In the fiducial models, we consider a simple scaling function for the fractional luminosity distance error, $d\log D = d\log D_0(1+\hat{z})$. In most cases, we adopt $d\log D_0=0.001$, and we discuss the effect from using smaller SNR events $d\log D_0 = 0.01$ with the catalog named \textit{\texttt{lum}}. We also adopt a scaling function for the angular uncertainty parameter, $\kappa(\hat{z}) = \kappa_0/(1+\hat{z})$. In the base case, we set $\kappa_0=10^4$, which corresponds to a typical scale of $d\theta_0=0.64$ deg. From our Fisher information error data hindsight, this is a reasonable value given the detector network we consider. To gauge the effect of worse localization error, in the fiducial model named \textit{\texttt{ang}}, we set $\kappa_0=10^3$, giving $d\theta_0 = 2.03$ deg. The choice of these target localization quality is motivated by Ref. \cite{ET_maggiore} (Figure 8) and Ref. \cite{TianGO} (Figure 3), where localization error can be below 1 deg in radius or less with next generation GW detector added to the network. For \textit{\texttt{ang}} only, we increase the simulated region to have a radius of 10 deg, to ensure we do not suffer from edge effects.

For realistic models, we compute measurement error using Fisher information. Firstly, we draw GW source intrinsic parameter samples. We consider the \texttt{POWER LAW + PEAK} model with a smoothing function at the low mass end. Distribution of the primary mass, $m_1$, is given as \cite{Abbott_pop}
\begin{equation}
    \pi(m_1) = [(1-\lambda_g)\mathcal{B}(m_1)+\lambda_g G(m_1)]S(m_1)\;,
\end{equation}
where $\mathcal{B}$ is the normalized power-law distribution with a slope of $-\alpha$ and a high-mass cutoff at $m_\mathrm{ max}$, and $G(m_1)$ is a normalized gaussian distribution with $(\mu_g,\sigma_g)$. The sigmoid smoothing function, $S(m_1)$, tapers the distribution at the low mass end with parameter $\delta_m$. The distribution of the secondary mass, $m_2$, follows a power law with a slope of $-\beta$. In this work, we adopt the parameters $\alpha = 3.78,\beta=0.81,\mu_g=32.27~M_\odot,\sigma_m=3.88~M_\odot,\lambda_g=0.03,\delta_m=2.5~M_\odot$, following Ref. \cite{Hang}, which are mostly identical to the inferred parameters in Ref. \cite{Abbott_gwtc3}. We assume the primary black hole mass is between $m_\mathrm{ min} = 6.5~M_\odot,m_\mathrm{ max}=112.5~M_\odot$. Using this distribution, we sequentially generate $(m_1,m_2)$ pairs. 

To compute GW signal-to-noise ratio (SNR) and parameter uncertainties, we assume a detector network of CE+LL+ET, a Cosmic Explorer \cite{CE2019} at current LIGO Hanford location, aLIGO Livingston and Einstein Telescope \cite{ET2011}. The noise power spectral density (PSD) is imported from the \texttt{psd} module from \texttt{PyCBC} \cite{pycbc}\footnote{Specifically, the adopted PSD versions are \texttt{CosmicExplorerP1600143}, \texttt{aLIGODesignSensitivityP1200087} and \texttt{EinsteinTelescopeP1600143} respectively.}. The detector antenna patterns, $f_p,f_c$, are obtained through \texttt{PyCBC.Detector.antenna\_pattern}\footnote{Specifically, with detectors \texttt{H1,L1} and \texttt{E1}.}. The plus, cross waveforms, $h_p,h_c$, are obtained via \texttt{PyCBC.get\_fd\_waveform} using approximant \texttt{IMRPhenomHM} \cite{IMRHM}. 

We compute uncertainties from Fisher matrix for each mass pair on a grid of parameter space, $(\hat{z},\hat{\Omega},\iota)$. The redshift grid is evenly spaced from $\hat{z}=0.01$ to $2$. The $\iota$ grid (source orbital inclination) is evenly spaced from $\iota=0.01$ to $\pi/2-0.01$. The angular position $\hat{\Omega}$ is the \texttt{healpy}\cite{healpy} grid pixel with \texttt{NSIDE}=4 (in total 192 equal-areal pixels). 

In principle, these samples can be fed into Monte Carlo integration for Hubble error calculation (Eqn. \eqref{ch5_eqn:dNdMdD}); in practice such a simulation is time-consuming, since both the catalog piece and the theoretical piece in the Fisher information must be recomputed from convolution between the galaxy catalog and the new uncertainties. In the current scope of this study, we use the data from only one mass pair sample $(m_1,m_2)=(8.28~M_\odot,7.42~M_\odot)$. We have checked that this mass pair is close to the peak probability in the mass distribution, thus may be used as a proxy to the expected population majority. 

The corresponding parameter uncertainty is a weighted average over sky location and binary inclination angles, $(\hat{\Omega},\hat{\iota})$. For sky coverage, we take a flat average. The threshold $\iota$ depends on selection criterion. In our realistic simulations, we examine using either the SNR or the angular error. We find that at each redshift, the relative change in the criterion compared to when $\iota=0$ demonstrates a universal pattern, and we define this averaged scaling function as (here we use SNR, $\rho$, as the example) 
\begin{equation}
    g(\iota) = \mathrm{interp1d}\left(\iota, \frac{1}{N_z}\sum_{i=0}^{N_z}\frac{\rho(\iota,z_i)}{\rho(0,z_i)}\right)\;.
\end{equation}
Specifically, we used linear interpolation from \texttt{scipy.interp1d} \cite{scipy}. At each redshift, the threshold $\iota$ can be computed from the inverse function, and we take the volume-weighted average as the nominal uncertainty, e.g.,
\begin{equation}
    \begin{split}
        \iota_\mathrm{thresh}(z) &= g^{-1}\left(\frac{\rho_\mathrm{thresh}}{\rho(\iota=0,z)}\right)\\
        \delta\overline{\theta} &= \frac{1}{N_\Omega}\sum_{i=1}^\mathrm{NPIX}\frac{\int_{\iota_\mathrm{thresh}}^{\pi/2}\delta\theta \sin\iota d\iota}{\int_{\iota_\mathrm{thresh}}^{\pi/2} \sin\iota d\iota}\;.
    \end{split}
\label{eqn_ithresh}
\end{equation}

To visualize the resulting rates model, we produce four model cases using $\rho_\mathrm{thresh}=100,200$ (\textit{\texttt{rho\_sm}}, \textit{\texttt{rho\_lg}}) and $\theta_\mathrm{thresh}=0.4~\mathrm{deg},1~\mathrm{deg}$ (\textit{\texttt{theta\_sm}}, \textit{\texttt{theta\_lg}}).

\subsubsection{GW completeness fraction}

As is seen in Eqn. \eqref{eqn:gw_compfrac}, GW completeness fraction provides the overall scaling for how many events we can expect to have for the Hubble measurement. In the fiducial models, we adopt a phenomenological approach and use a sigmoid function,
\begin{equation}
    f_\mathrm{gw} = \frac{1}{1+\exp((\hat{z}-\mu_\mathrm{cf})/\sigma_\mathrm{cf})}\;,
\label{eqn:sigmoid}
\end{equation}
and we hold $\sigma_\mathrm{cf}=0.15$, which controls the decay rate of the completeness fraction.

In models with simulated measurement uncertainty, the GW completeness fraction can be directly obtained by enforcing selection criterion. Assuming that the orbital axis of the binary is uniformly distributed over the sphere, the completeness fraction is given by $f_\mathrm{gw}=1-\cos\iota_\mathrm{thresh}$ (see Eqn. \eqref{eqn_ithresh}). This setup brings an important difference from the fiducial models. In the fiducial cases, the rates model is only affected by an overall scaling upon completeness fraction changes, since we assume source selection is independent from measurement quality. In the realistic case, the completeness fraction is closely related to averaged measurement quality; as the selected GW sources decrease in number, the selection is more stringent, and individual event quality improves.

\begin{figure*}[hbt!]
\centering
\includegraphics[width=\textwidth]{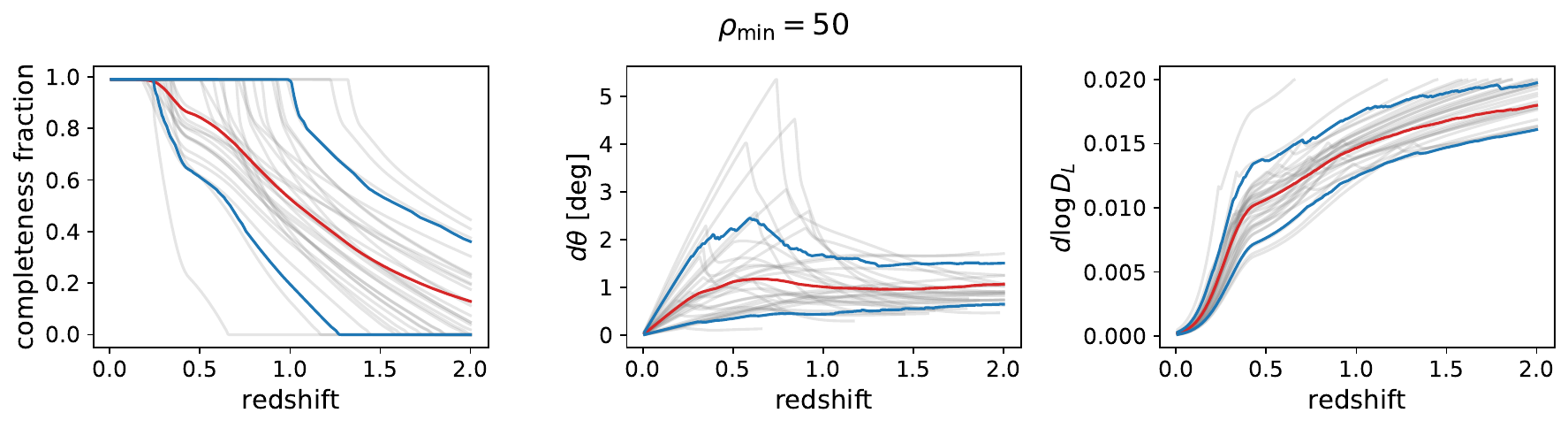}
\includegraphics[width=\textwidth]{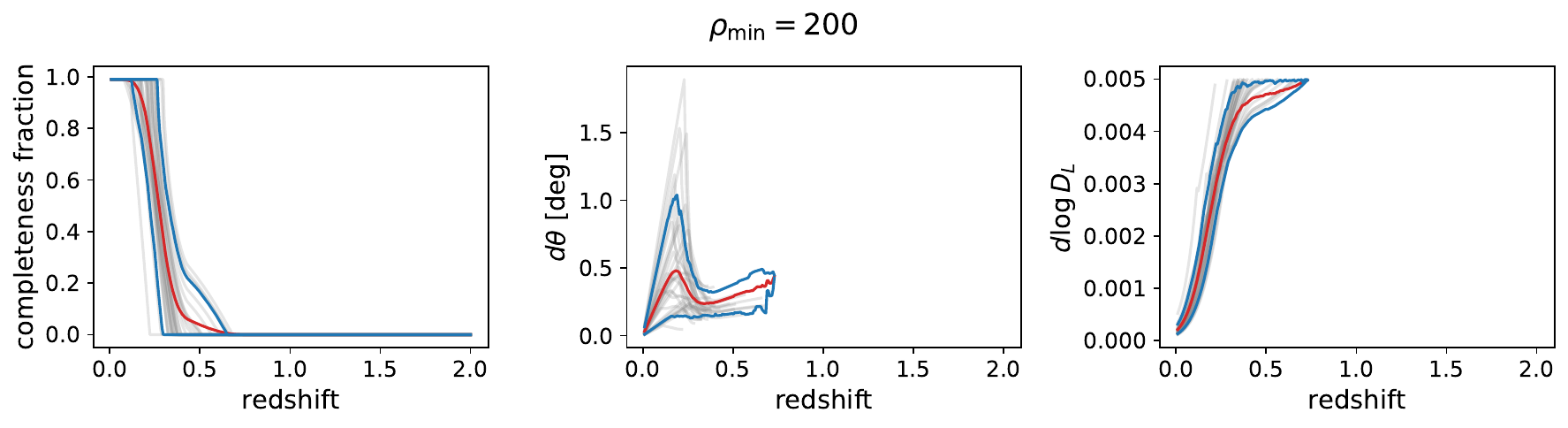}
\includegraphics[width=\textwidth]{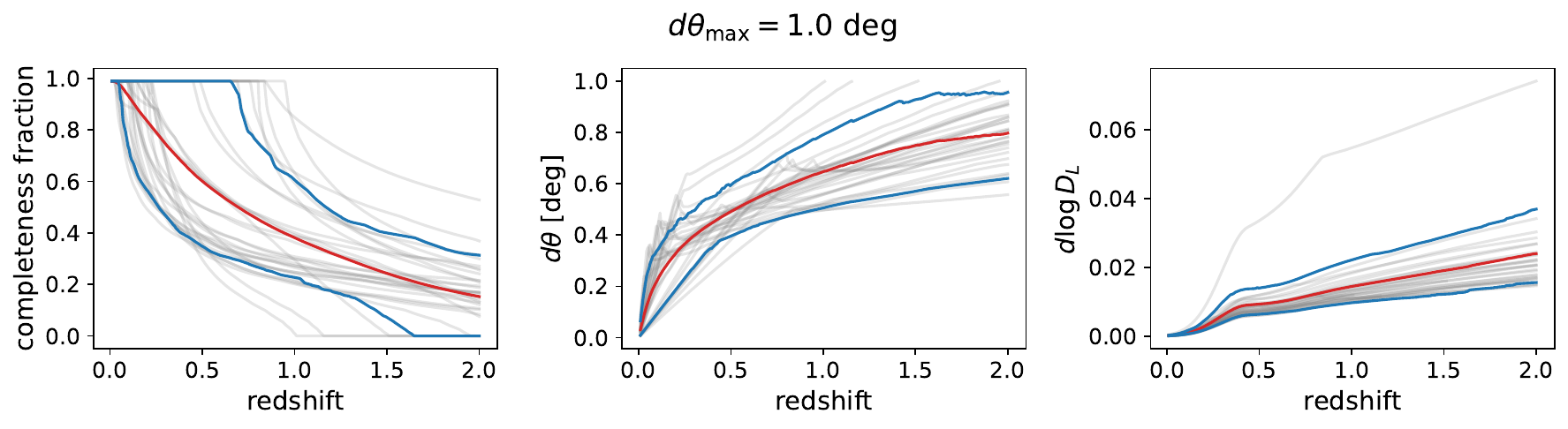}
\includegraphics[width=\textwidth]{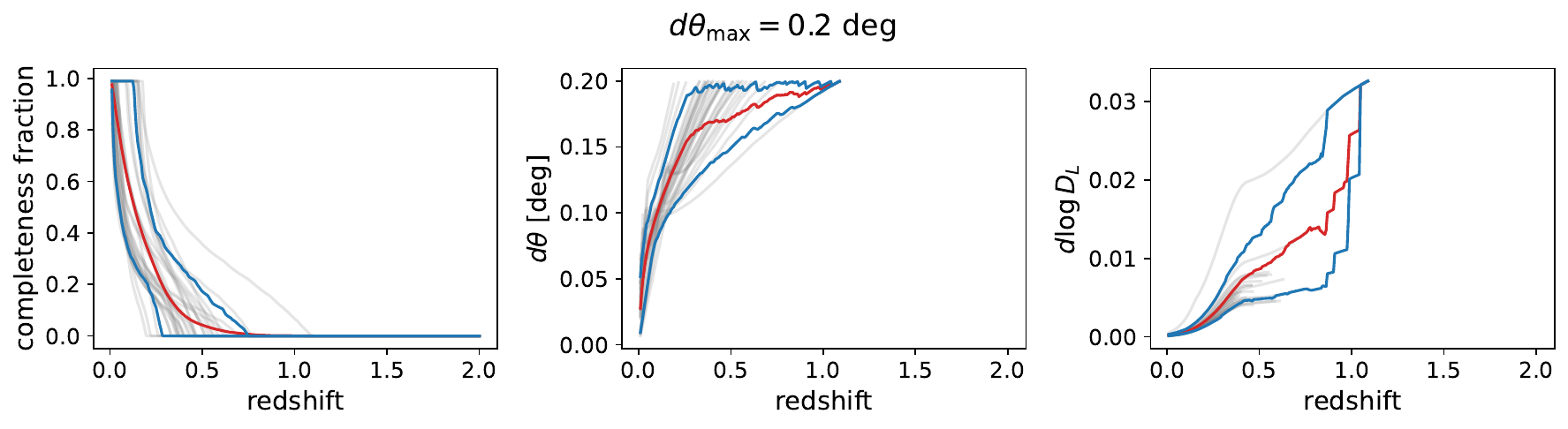}
\caption[Example completeness fraction and GW measurement uncertainties.]{Example completeness fraction and GW measurement uncertainties. From \textit{Top} to \textit{Bottom} panels, the selection criteria are for an SNR of 50, 200, and a maximum angular uncertainty (radius) of 1, 0.2 deg. In each panel, the light gray traces show results for randomly selected 30 sky pixels. The red trace shows the pixel-average that is adopted in the final Fisher calculation. The blue bracket shows the 80 percentile. In the uncertainty panels, the traces end when no sky pixels contain above-threshold sources.}
\label{fig_deltas}
\end{figure*}

\begin{figure*}[hbt!]
\centering
\includegraphics[width=\textwidth]{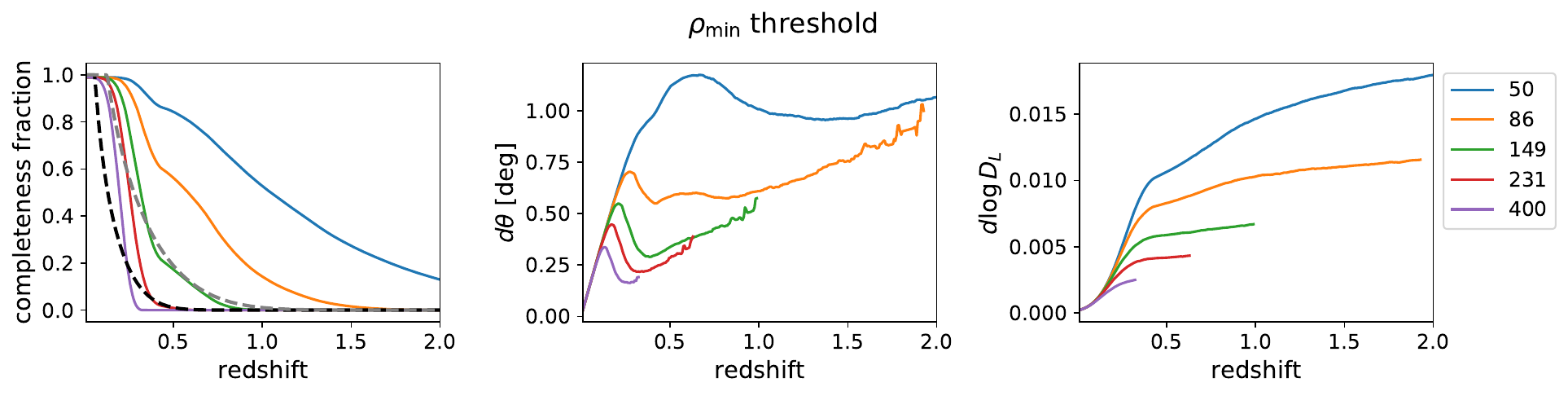}
\includegraphics[width=\textwidth]{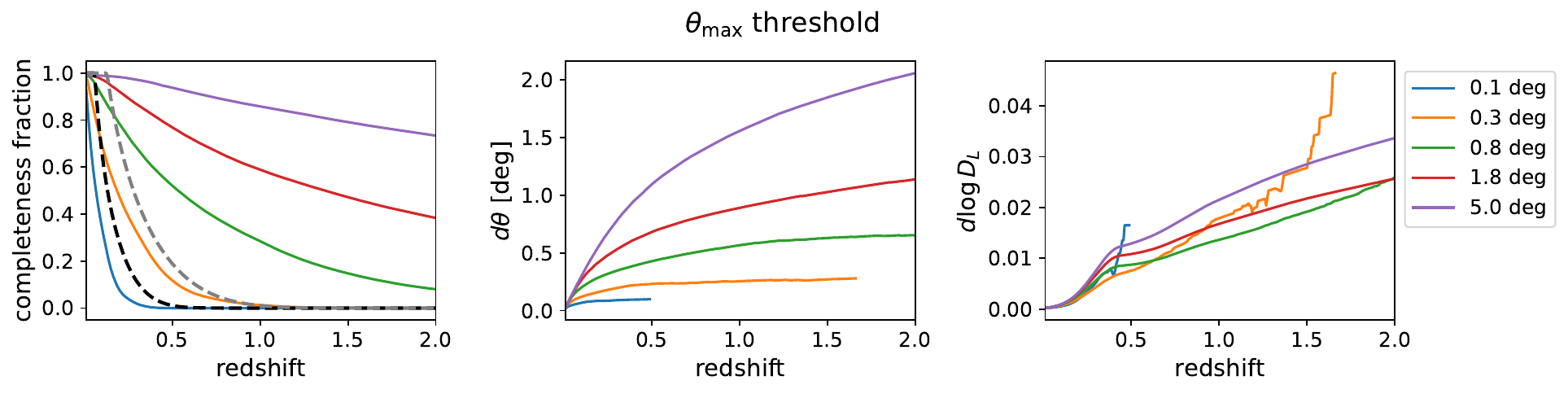}
\caption{Volume-averaged completeness fraction and GW measurement uncertainties. The \textit{Top} panel shows the SNR criterion and the \textit{Bottom} panel shows the angular resolution cap. The galaxy completeness fraction with magnitude limit of 20.175 and 22 are shown with dashed black and gray traces, respectively.}
\label{fig_delta_scan}
\end{figure*}

In Figure \ref{fig_deltas}, we show the weighted completeness fraction, angular uncertainty and luminosity distance uncertainty given several selection criteria. We show results for a random selection of 30 sky pixels (gray trace), and show the adopted mean value (red trace) and the 80 percentile (blue trace). In Figure \ref{fig_delta_scan}, we show the adopted mean value for a range of thresholds that we scan over in the final simulation. In panels for uncertainties, the traces end when no sky pixels contain qualifying sources. We observe peculiar behavior in, e.g. the angular uncertainty when using SNR as the threshold, shown as a bump in intermediate redshift values. This is due to the difference in uncertainty growth rate for localization and SNR. By excluding the inclination values with sub-threshold SNR, the remaining \textit{selected} GW sources with better measurement quality may have a smaller weighted angular uncertainty. We also show the galaxy completeness fraction with a magnitude limit of 19.5 and 20.175 in dashed black and gray lines for comparison. 

\subsubsection{GW event number density}
The GW number density is constrained from observation; based on events up to GW170104, the local BBH merger rate is constrainted to be $103^{+110}_{-63}~\mathrm{Gpc}^{-3}\mathrm{yr}^{-1}$ \cite{GW170104}. With GWTC2, the local merger rate for BBH is updated to $23.9^{+14.3}_{-8.6}~\mathrm{Gpc}^{-3}\mathrm{yr}^{-1}$ \cite{gwtc2_localrates}. We assume that this rate is a constant in redshift and across galaxies, due to the lack of relevant observational constraints. In our fiducial models, we found that the GW uncertainties can be more conservative compared to actual Fisher matrix results, resulting in a more pessimistic assessment on susceptibility to bias. To better showcase different effects, we artificially raise the GW event rates to $239~\mathrm{Gpc}^{-3}\mathrm{yr}^{-1}$ for most fiducial models. The realistic case is shown with model named \textit{\texttt{realistic\_bbh\_rate}}. In the realistic models, we found that using realistic BBH merger rates already produces simulation results with sufficient clarify.

Based on variations of the above model parameters, we present eight fiducial models and four models with realistic GW uncertainties, summarized in Table.~\ref{tables_fid_catalog_params} and \ref{tables_fisher_catalog_params}. All models except \textit{\texttt{ang}} share the same simulated volume. Realistic models have identical galaxy parameters and GW occupation rate as \textit{\texttt{realistic\_bbh\_rate}}.

\begin{sidewaystable}[hbt]
\centering
\renewcommand{\arraystretch}{1.25}
\begin{tabular} { | C{11em} | C{5em} | C{5em} | C{5em} | C{3em} | C{5em} | C{5em} | C{5em} | C{10em} |}
 \hline
 catalog name & $\sigma_g$ & $m_g$ & $d\log \hat{D}$ & $\hat{\kappa}$ & equiv. $\theta_0$ [deg] & $f_\mathrm{g, red}$ & $r_\mathrm{obs,GW}$ $ [\mathrm{Gpc}^{-3}\mathrm{yr}^{-1}]$\\
 \hline\hline
 \textit{\texttt{optimistic}} & 0.003 & 20.175 & 0.001 & $10^4$ & 0.8 & 1. & 239  \\
\hline
\textit{\texttt{realistic\_bbh\_rate}} & 0.003 & 20.175 & 0.001 & $10^4$ & 0.8& 1. & 23.9  \\
\hline
\textit{\texttt{bright}} & 0.003 & 19.5 & 0.001 & $10^4$ & 0.8& 1. & 239  \\
\hline
\textit{\texttt{photo}} & 0.01 & 20.175 & 0.001 & $10^4$ & 0.8& 1. & 239  \\
\hline
\textit{\texttt{ang}} & 0.003 & 20.175 & 0.001 & $10^3$ & 2.3 & 1. & 239  \\
\hline
\textit{\texttt{lum}} & 0.003 & 20.175 & 0.01 & $10^4$ & 0.8& 1. & 239  \\
\hline
\textit{\texttt{sparse}} & 0.003 & 20.175 & 0.001 & $10^4$ & 0.8& 0.1 & 239  \\
\hline
\textit{\texttt{deepz}} & 0.003 & 22 & 0.001 & $10^4$ & 0.8& 1. & 239  \\
\hline
\end{tabular}
\caption{Table of fiducial model catalogs parameters. From left to right, the columns represent catalog name, galaxy redshift uncertainty $\sigma_g$, galaxy survey magnitude limit $m_g$, fractional uncertainty in GW event luminosity distance $d\log \hat{D}$, GW angular uncertainty parameter coefficient $\hat{\kappa}$, the equivalent angular uncertainty at $z=0$ , $\theta_0$, galaxy number density scalar $f_\mathrm{g, red}$ and GW number density $r_\mathrm{obs,GW}$. See text for detailed explanation of parameters.}
\label{tables_fid_catalog_params}
\end{sidewaystable}

\begin{table*}[hbt]
\centering
\renewcommand{\arraystretch}{1.25}
\begin{tabular} { | C{7em} | C{12em} | C{8em} |}
 \hline
 catalog name & threshold type & threshold value \\
 \hline\hline
 \textit{\texttt{theta\_sm}} & angular uncertainty & 0.4 deg \\
\hline
 \textit{\texttt{theta\_lg}} & angular uncertainty & 1. deg \\
\hline
 \textit{\texttt{rho\_sm}} & SNR & 100 \\
\hline
 \textit{\texttt{rho\_lg}} & SNR & 200 \\
\hline
\end{tabular}
\caption{Table of model catalog parameters using realistic GW measurement uncertainty from Fisher information calculations. From left to right, the columns represent catalog name, source selection threshold type and the corresponding threshold value. The galaxy catalog parameters and number density scaling is the same as the fiducial model \textit{\texttt{realistic\_bbh\_rate}}.}
\label{tables_fisher_catalog_params}
\end{table*}

\section{Simulation Result Analysis}
\label{chap:analysis}
\begin{figure*}[htb!]
    \centering
    \includegraphics[width=0.8\textwidth]{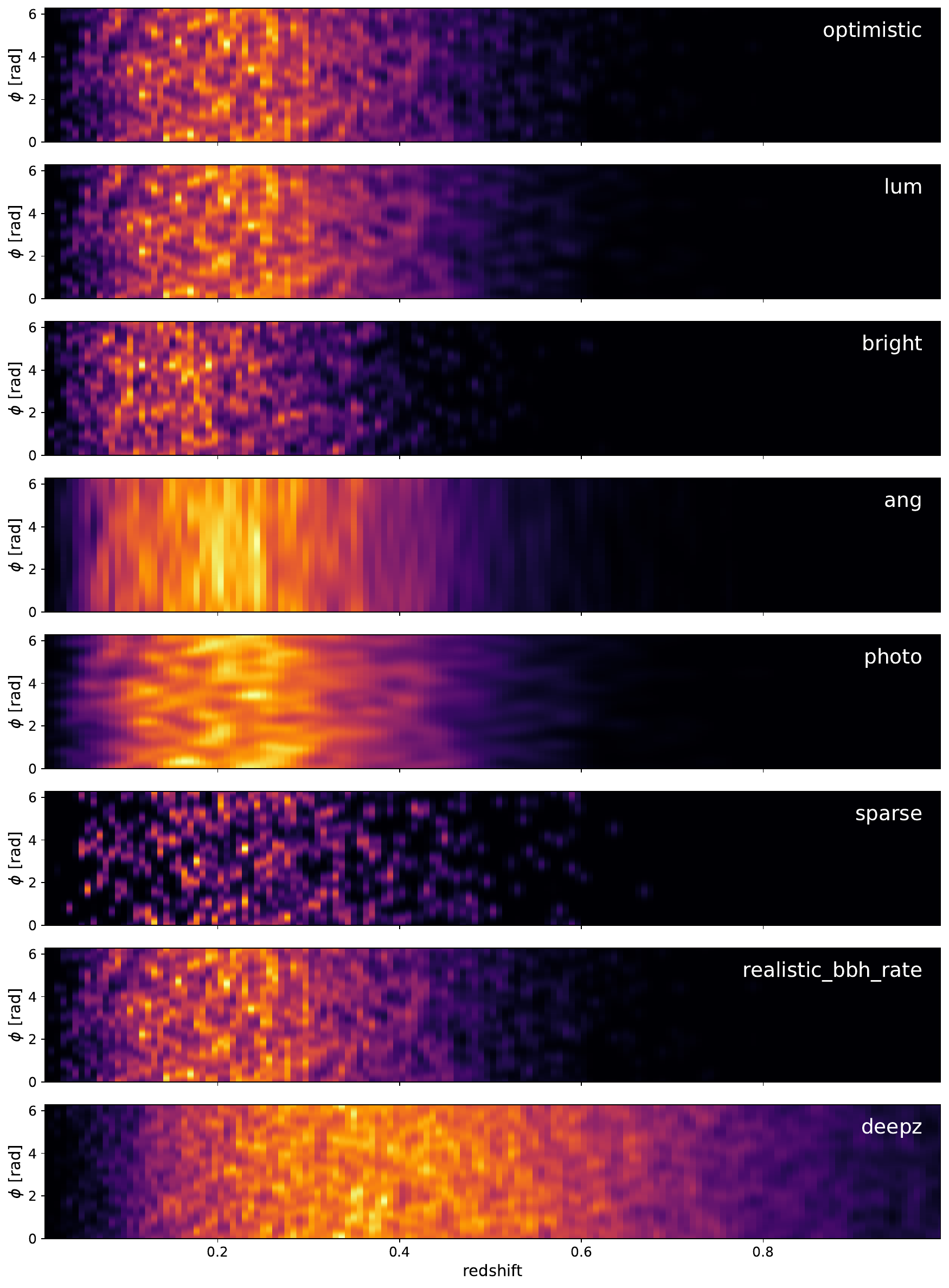}
    \caption{Slice of the expected GW event rate due to galaxy catalog in the simulated volume at $\theta=3.3$ deg. Each panel represents a fiducial catalog with the catalog name shown on the upper right corner. See text and Table~\ref{tables_fid_catalog_params} for model parameters. Note that each panel has an independent color scale.}
    \label{fig_fid_all_zslice}
\end{figure*}

\begin{figure*}
    \centering
    \includegraphics[width=0.8\textwidth]{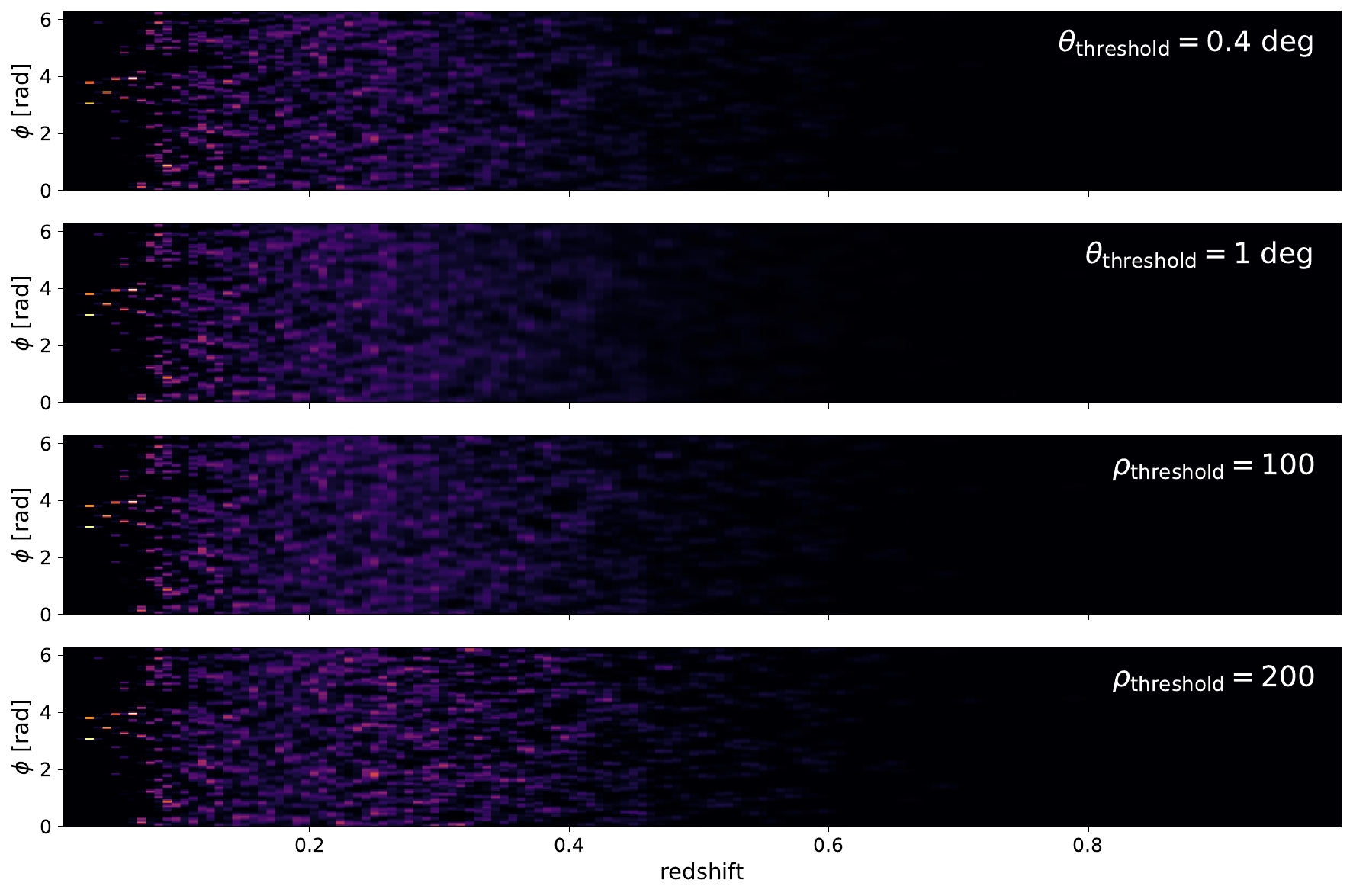}
    \caption{Slice of the expected GW event rate due to galaxy catalog in the simulated volume at $\theta=3.3$ deg. Each panel represents a catalog with the catalog name shown on the upper right corner. See text and Table~\ref{tables_fid_catalog_params} for model parameters. Note that each panel has an independent color scale.}
    \label{fig_fisher_all_zslice}
\end{figure*}

\begin{figure*}
    \centering
    \includegraphics[width=0.6\textwidth]{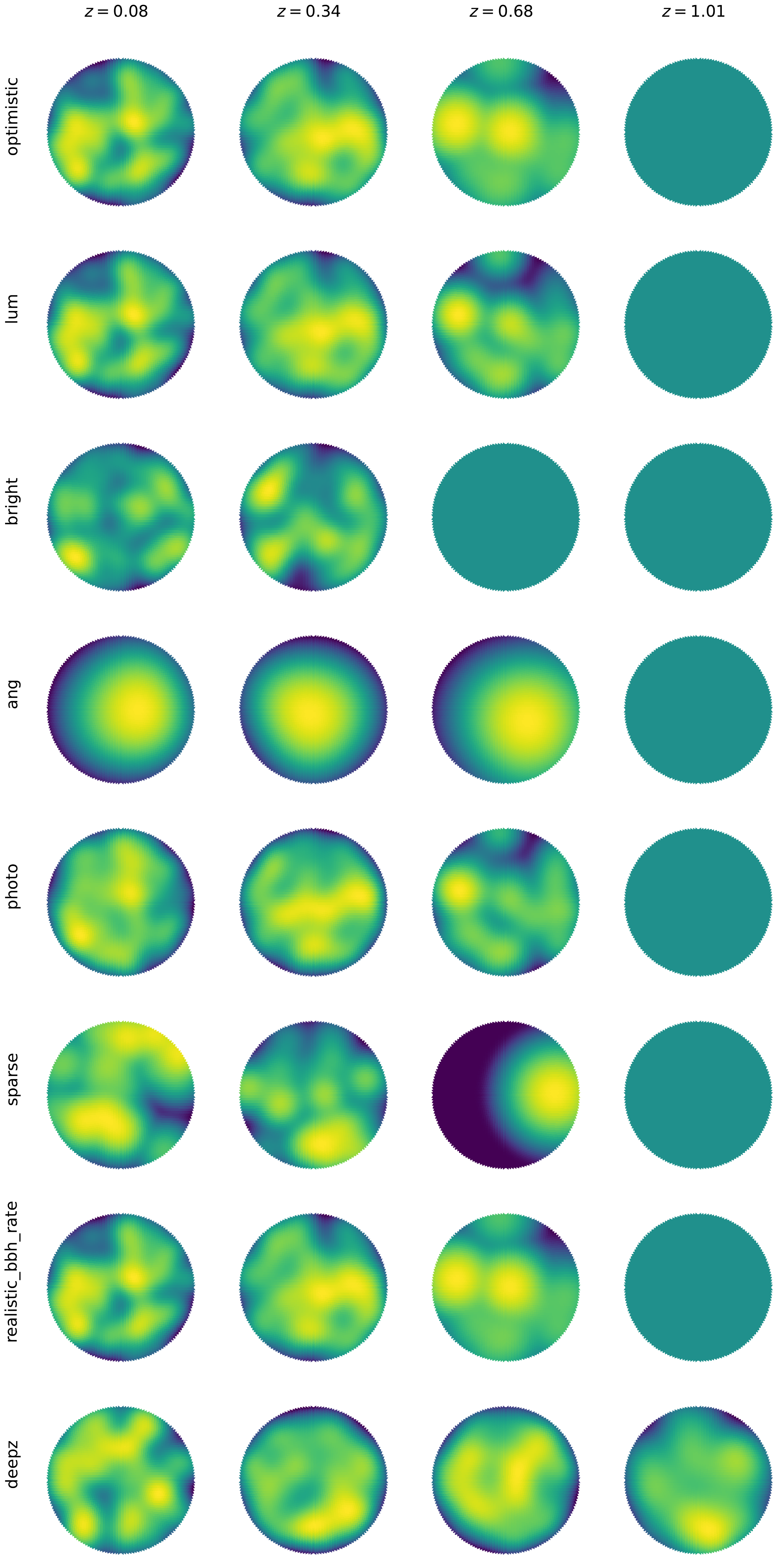}
    \caption{Angular cross section of the expected GW event rate due to galaxy catalogs. The galaxy catalog is convolved with fiducial values of GW measurement uncertainty. From \textit{Top} to \textit{Bottom}, catalog names are labeled on the left. From \textit{Left} to \textit{Right}, the redshift is $(0.08,0.34,0.68,1.01)$. Note that each panel has an independent color scale. See text and Table~\ref{tables_fid_catalog_params} for simulation parameters.}
    \label{fig_fid_all_ang}
\end{figure*}

\begin{figure*}
    \centering
    \includegraphics[width=0.8\textwidth]{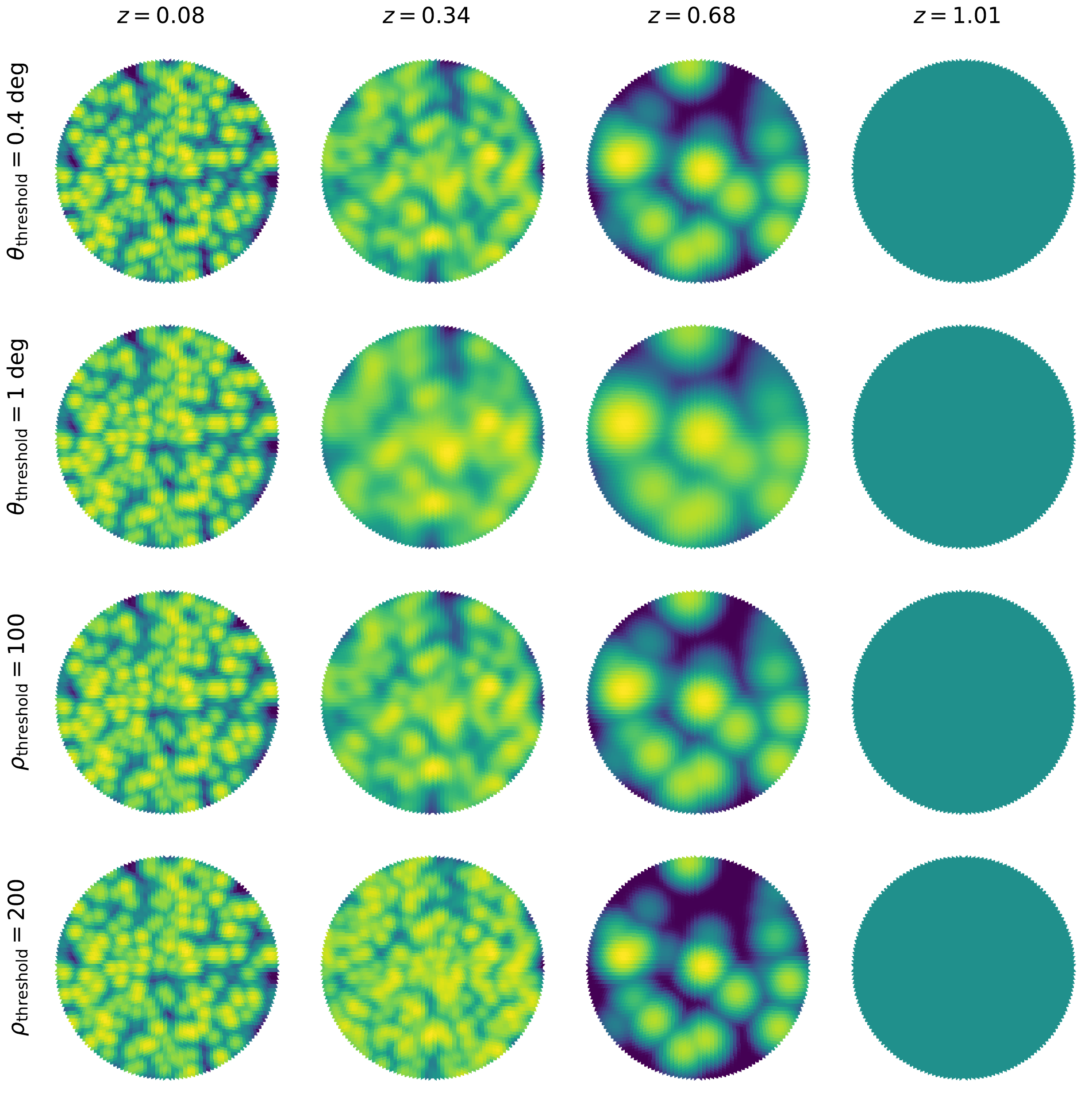}
    \caption{Angular cross section for simulated catalogs with realistic GW uncertainties at different redshifts. The GW event selection criterion is noted on the left. From \textit{Left} to \textit{Right}, the redshift is $(0.08,0.34,0.68,1.01)$. See text and Table~\ref{tables_fisher_catalog_params} for simulation parameters. Note that each panel has an independent color scale.}
    \label{fig_fisher_all_ang}
\end{figure*}

In this section, we analyze our simulation results and discuss implication to the GW dark siren Hubble measurement. In Figure \ref{fig_fid_all_ang} and Figure \ref{fig_fisher_all_ang}, we show angular cross section of the catalog-piece rates model, $r_\mathrm{cat}(\hat{D})$, at different redshifts. In Figure \ref{fig_fid_all_zslice} and Figure \ref{fig_fisher_all_zslice}, we unwrap $r_\mathrm{cat}(\hat{D})$ at $\theta=3.3$ deg to show one slice along the redshift direction. All panels in these four figures have independent color scale. 

In Figure \ref{fig_fid_all_zslice}, we analyze the catalog-piece rate model by comparing to the base case, \textit{\texttt{optimistic}}. The models \textit{\texttt{lum}}, \textit{\texttt{ang}} and \textit{\texttt{photo}} show rate changes with poorer measurement sensitivity. In the redshift direction, while GW fractional errors in \textit{\texttt{lum}} increase an order of magnitude and the redshift error in \textit{\texttt{photo}} increases only by a factor of 3, the redshift-direction blurring effect is much more apparent in \textit{\texttt{photo}}. In the angular direction, the GW angular uncertainty in \textit{\texttt{ang}} increases only by $24\%$, yet the angular direction rates becomes significantly flatter. This contrast suggests that the redshift uncertainty is dominated by galaxy catalog errors, and the angular direction is dominated by GW localization. 

Comparing \textit{\texttt{sparse}} to \textit{\texttt{optimistic}}, we observe that the rates model becomes more sharply peaked. In \textit{\texttt{deepz}}, the catalog redshift reach is much deeper, although the higher redshift section rates model becomes less spiky due to larger redshift error and more galaxies blending together.

% By comparing Figure \ref{fig_fid_all_ang} (\textit{\texttt{realistic\_bbh\_rate}} has identical galaxy and overall event rate parameter to realistic models) and Figure \ref{fig_fisher_all_ang}, we observe much sharper rates in realistic models especially at smaller redshift. This suggests that the fiducial models This suggests that the fiducial models overestimate GW measurement error in the context of this detector network. Nonetheless, we emphasize that models should be compared mainly within the same broad group to control variables.

\begin{figure}[hbt]
    \centering
    \includegraphics[width=\columnwidth]{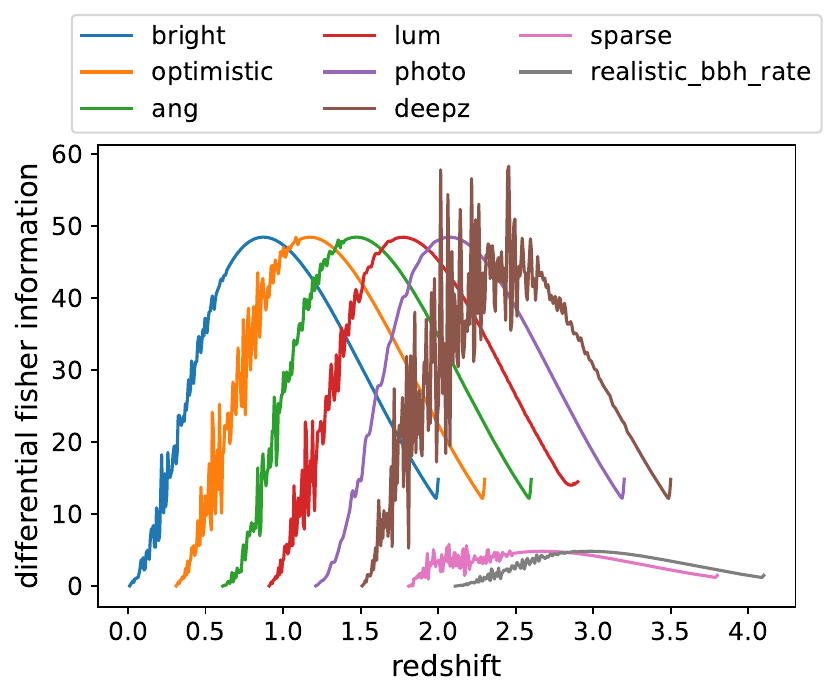}
    \caption{Full differential Fisher information for various fiducial models. The GW completeness fraction has not been applied. For presentation clarity, all but the leftmost trace in each panel have been shifted to larger redshift values.}
    \label{fig_fid_fisher_info}
\end{figure}

\begin{figure*}[hbt!]
\centering
\includegraphics[width=0.8\textwidth]{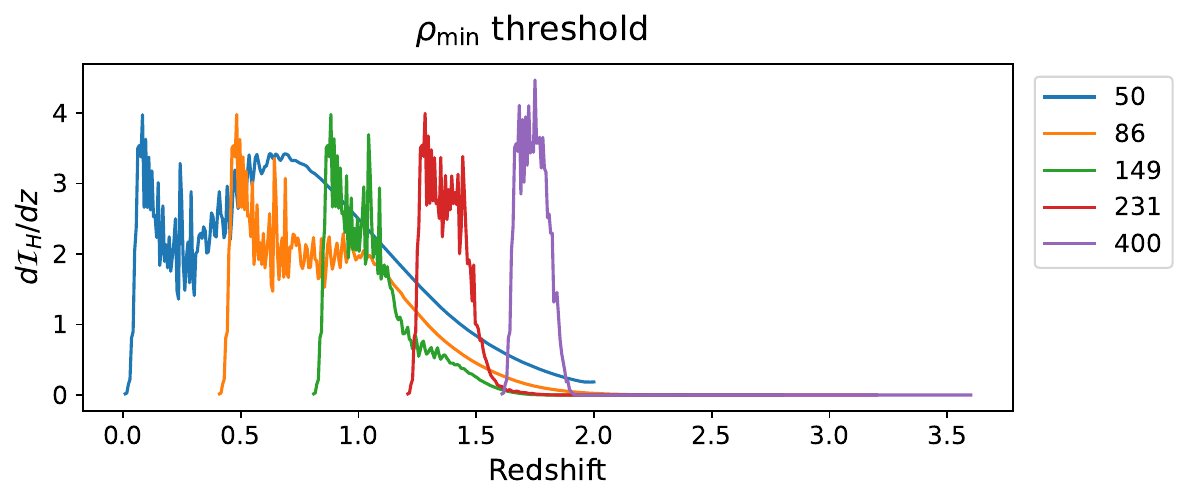}
\includegraphics[width=0.8\textwidth]{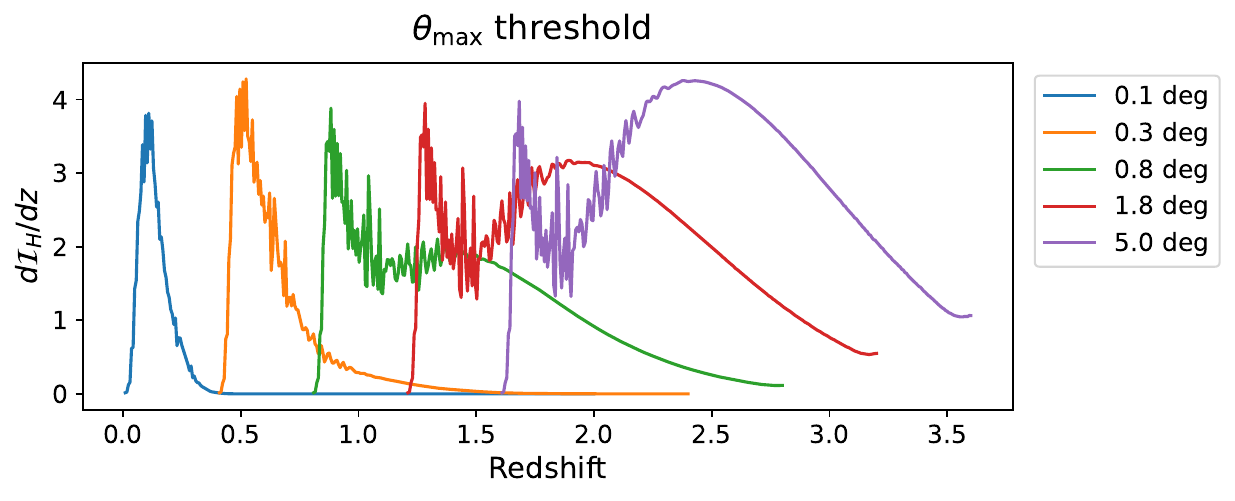}
\caption{Differential fisher information for realistic models given different selection criteria. The GW completeness fraction is applied. In all except the leftmost traces, the redshift coordinate has been artificially shifted to the right for visual clarify.}
\label{fig_consis_IH}
\end{figure*}

In Figure \ref{fig_fid_fisher_info}, we plot the differential Fisher information for various fiducial models prior to scaling with GW completeness fraction. For clarify, all traces except the leftmost one have been shifted to the right. As expected, the overall scale of the Fisher information is similar for models with the same galaxy and GW number density, since galaxy completeness and measurement quality only contribute to the ratio between the catalog piece and the theoretical piece, as well as the model smoothness. The smoothing degree and spiky region location are consistent with the event rate angular and redshift direction cross section (e.g. comparing \textit{\texttt{photo}} and \textit{\texttt{deepz}}).

In Figure \ref{fig_consis_IH}, we show the differential Fisher information for realistic models using either SNR or localization as the selection criterion. Since in these models the GW completeness fraction is fixed given a selection criterion, the differential Fisher information shown has already been scaled with $f_\mathrm{gw}$. In both panels, we observe that more stringent selection criterion produces taller peaks at small redshift, but the redshift reach also drastically reduces. The final Fisher information depends on the competition between these two factors. 

In both groups, we observe that the Fisher information peak is misaligned with the galaxy distribution, which peaks around $z=2$. While at higher redshift the expected events are more numerous, their contribution to constraining the Hubble constant decreases due to their flatter profile. This phenomenon becomes more dramatic in realistic models, where the Fisher information shows an additional peak at very low redshift. In combination with Figure \ref{fig_fisher_all_zslice}, this demonstrates that low-measurement-error events at small redshift contribute significantly to constraining the Hubble constant albeit the disadvantage in number. 

\section{Galaxy Model Error Tolerance}
\label{sec:beta_analysis}

As is shown in Eqn. \eqref{eqn:bias}, bias from incorrect galaxy model assumption is sensitive to the form of rates difference. In Ref. \cite{Hang}, the considered galaxy model correction are local overdensities of galaxies, and they are modeled as individual gaussian bumps in the rates model. In Ref. \cite{dalang}, the authors consider galaxy clustering and generate test catalogs given real space correlation function, rather than having uniformly distributed Poisson samples; in this case the model error would be the deviation from these large scale structures. In this work, we consider the case where the assumed galaxy mass function has a different redshift dependence from the true distribution,
\begin{equation}
    r = r_\mathrm{ true}(1+\hat{z})^\beta\;.
\end{equation}
In this case, the catalog piece $r_\mathrm{ cat}$ remains unchanged while the supplemented theoretical rate is multiplied with $(1+\hat{z})^\beta$. The corresponding error rate term is then given by $\delta r = \left[1-(1+\hat{z})^\beta\right]r_\mathrm{ true}$. The statistical uncertainty and the bias is computed according to Eqn. \eqref{eqn:dHs} and Eqn. \eqref{eqn:bias}.

Intuitively, mitigating potential bias and reducing statistical uncertainties of Hubble measurement place different demands on the GW events used for inference. On one hand, the potential bias does not benefit from more events; it requires that the incorrect rate piece plays less role in the Fisher information. Therefore, reducing bias requires placing emphasis on close-by events where galaxy catalog is highly complete. On the other hand, the statistical uncertainty depends only on the total Fisher information, which benefits greatly from including high-redshift events occupying greater volume. Therefore, considering both pieces is essential for balancing the dark siren precision and accuracy. In the following test, we investigate that, given a fixed total error budget, what the maximum tolerated redshift evolution parameter $\beta$ is under various GW event selection. 

% Since galaxy completeness reduces typically much earlier than the expected peak of galaxy mass function, reducing bias requires placing emphasis on the close-by well-resolved events. 

\begin{figure}[hbt!]
    \centering
    \includegraphics[width=\columnwidth]{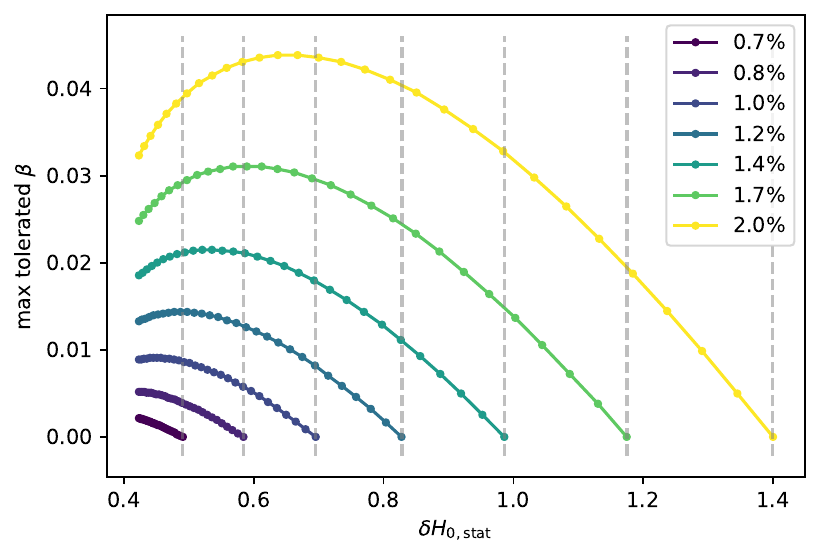}
    \caption{Maximum galaxy model error tolerance given various total error budget using the fiducial models \textit{\texttt{realistic\_bbh\_rate}}. The target total error budget is geometrically distributed between $0.7\%$ and $2\%$. The horizontal axis has a unit of km/s/Mpc. }
    \label{fig_beta_threshold_realisticbbh}
\end{figure}

In Figure \ref{fig_beta_threshold_realisticbbh}, we study fiducial models. We vary the GW selection completeness fraction, i.e., $\mu_\mathrm{cf}$ in Eqn. \eqref{eqn:sigmoid}, and search for the maximum galaxy mass function redshift bias $\beta$ for a fixed total error budget. This figure shows the result for the model \textit{\texttt{realistic\_bbh\_rate}}. The searched total error budget values are geometrically distributed from 0.7\% to 2\%, and we mark them with vertical dashed lines. The right end of the traces marks either when the statistical error saturates the total budget, or when $\mu_\mathrm{cf}=0.01$ at the edge of our simulation volume. The left end marks either when the statistical error is $10\%$ of the total budget, or when $\mu_\mathrm{cf}=1.8$, close to the maximum simulated redshift. As the total error budget changes, we observe the development of a sweet spot; prior to the sweet spot, i.e., the right-hand-side of the figure, the improvement in statistical error is faster than the increasing contribution from $r_\mathrm{theo}$, which leaves greater room for theoretical model induced bias. However, as we keep adding deeper sources, the growth of Fisher information is suppressed due to the smoothing of the rates model, and is soon outpaced by the bias brought by the large theoretical population. This observation suggests that, given the same error budget, if we would like to prioritize robustness against variations in astrophysical models, it can be desirable to apply additional filtering and disregard distant events.

\begin{figure}[hbt!]
    \centering
    \includegraphics[width=\columnwidth]{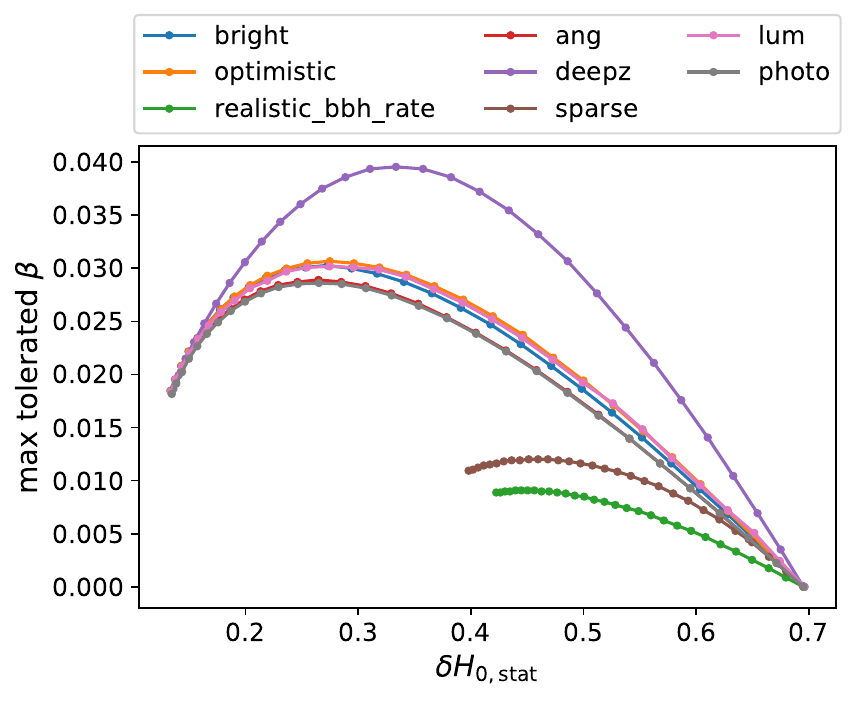}
    \includegraphics[width=\columnwidth]{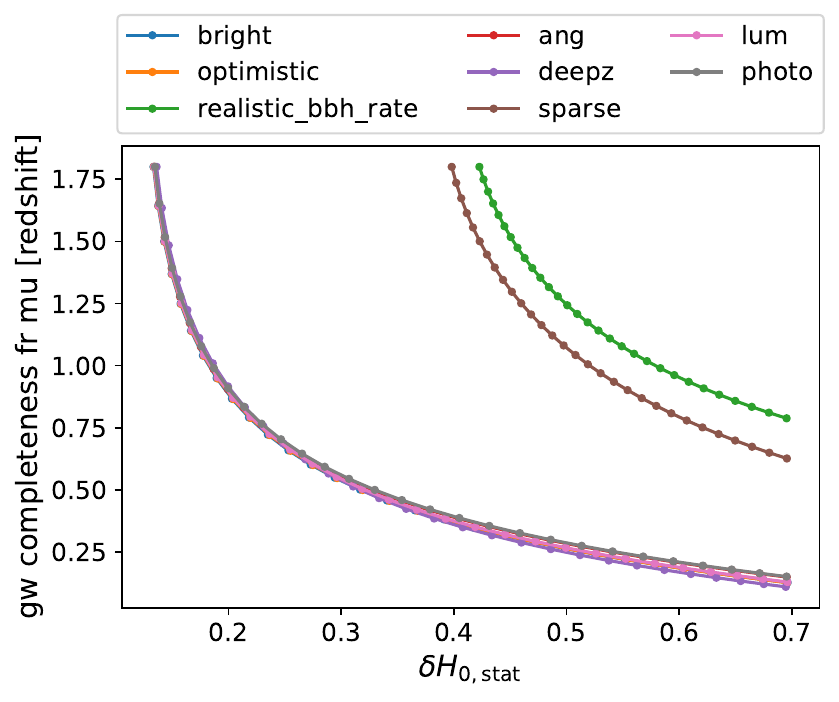}
    \caption{Maximum galaxy model error tolerance with $1\%$ total error budget for fiducial simulation models. The \textit{Left} panel shows the maximum $\beta$, and the \textit{Right} panel shows the corresponding $\mu_\mathrm{cf}$ in GW completeness fraction. See text for ending point definition. The horizontal axis has a unit of km/s/Mpc.}
    \label{fig_fid_percent}
\end{figure}

\begin{table*}[hbt]
\centering
\renewcommand{\arraystretch}{1.25}
\begin{tabular} { | C{7em} | C{7em} | C{5em} |C{5em} |C{5em} |}
 \hline
 catalog & \textit{\texttt{optimistic}} & \textit{\texttt{bright}} & \textit{\texttt{sparse}} & \textit{\texttt{deepz}}\\
 \hline\hline
 $N_\mathrm{gal}$ & 7047 & 3392 & 684 & 45251\\
\hline
\end{tabular}
\caption{Number of simulated galaxies in each model. Note that models with the same galaxy completeness fraction and number density share the same galaxy catalog.}
\label{tables_ngal}
\end{table*}

In Figure \ref{fig_fid_percent}, we show the galaxy model error tolerance given different fiducial model parameter assumptions for a 1\% total error in the \textit{Top} panel. In the \textit{Bottom} panel, we show the corresponding GW completeness fraction parameter $\mu_\mathrm{cf}$. The overlay of the traces demonstrates that given a similar galaxy catalog, small variations of measurement uncertainty does not significantly vary the overall Fisher information, and the same statistical uncertainty is achieved with similar GW observational depth. By comparing the error tolerance, however, we observe that \textit{\texttt{sparse}} shows significant improvement over \textit{\texttt{realistic\_bbh\_rate}}, even though they both represent a density reduction of a factor of 10. This is due to the fact that the Fisher information is highly sensitive to the \textit{derivative} of the rates model; changing the GW density only applies an overall scaling, but increasing galaxy density leads to more blending between neighboring galaxy probability region and ``flattens out'' the derivative. In the completeness figure, we also see that \textit{\texttt{sparse}} achieves the same level of $\delta H_\mathrm{0,stat}$ at smaller redshift, further supporting the derivative argument.

\begin{figure}[hbt!]
    \centering
    \includegraphics[width=\columnwidth]{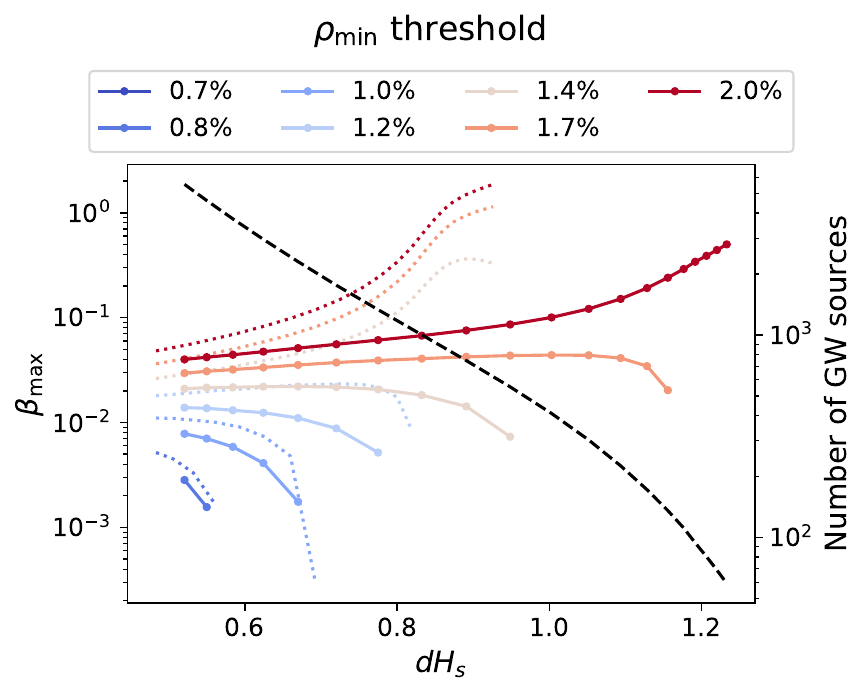}
    \includegraphics[width=\columnwidth]{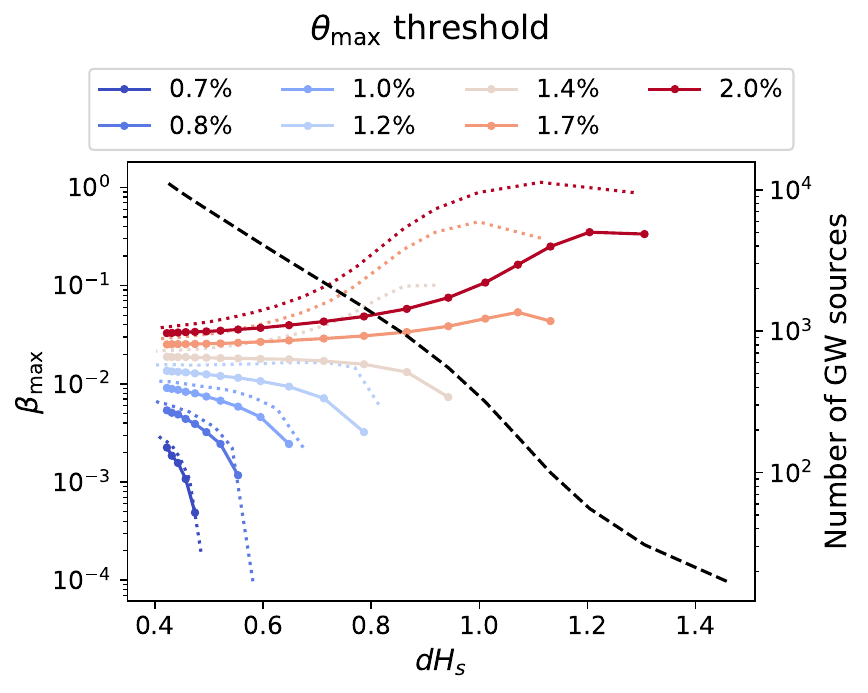}
    \caption{Maximum galaxy model error tolerance given various total error budget. The \textit{Top} panel shows SNR selection and the \textit{Bottom} shows the angular resolution criterion. In each panel, the dotted lines shows model error tolerance with clustered galaxy catalog, using the same color scheme. The target total error budget is geometrically distributed between $0.7\%$ and $2\%$. The right vertical axis shows the expected number of GW sources for each threshold. In the \textit{Top} panel, no selection criterion gives statistical uncertainty at $0.7\%~H_0$, and the corresponding trace is absent. The horizontal axis has a unit of km/s/Mpc.}
    \label{fig_beta_threshold_consis}
\end{figure}

In Figure \ref{fig_beta_threshold_consis}, we show tolerance to $\beta$ using realistic GW uncertainties as we apply different source selection criteria. In the \textit{Top} panel, the minimum SNR $\rho_\mathrm{min}$ is geometrically distributed from 50 to 400, and in the \textit{Bottom} panel, the maximum angular localization radius $\theta_\mathrm{max}$ ranges from 0.2 deg to 5 deg following a geometric distribution. The black dashed line corresponds to the right vertical axis and shows the number of expected GW sources under each threshold condition. For the SNR selection, none of our searched value gives a statistical uncertainty below 0.7\%, and the corresponding trace is absent. 

For stringent total error budget, e.g. 0.7\% and 0.8\%, the behavior is generally similar to that for fiducial models, where the reduction in statistical error brings advantage to model error tolerance. With larger error budget, however, we observe a qualitative difference; rather than developing selection sweet spots, the error tolerance increases exponentially with increasing statistical uncertainty. In this case, it becomes more advantageous to apply stringent GW source selection and pick out close-by and well-localized sources. Since the galaxy catalog is highly complete at these distances (see e.g., Figure \ref{fig_delta_scan}), the inference is very robust to galaxy mass function assumptions. As an order-of-magnitude estimate, the GW uncertainty comoving volume for a target angular resolution of 0.2 deg and a redshift uncertainty of 0.003 (i.e. spectroscopic) at $z=0.2$, is $\sim 300 ~\mathrm{Mpc}^3$. Using the reference galaxy density of $0.002 ~\mathrm{Mpc}^{-3}$, this gives $<1$ galaxy in this volume, on average. Therefore, the dark sirens have effectively become bright sirens in this case. 

With the dotted lines in Figure \ref{fig_beta_threshold_consis}, we perform the same error threshold calculation but with a clustered galaxy catalog generated from MICE grand challenge. We note that in all test cases, the error tolerance increases when considering clustered galaxies. The improvement is especially dramatic in the ``effective bright siren'' scenario; in SNR-thresholding case, the error tolerance can improve over an order of magnitude with 2\% total error target. This result is analogous to the improvement of the fiducial \texttt{sparse} model over \texttt{realistic\_bbh\_rate}; increasing ``spikiness'' of the GW rates model improves the constraint on the Hubble constant. 

We now discuss the expected GW event number. From the two panels in Figure \ref{fig_beta_threshold_consis}, roughly 50 GW events can constrain the Hubble constant value to below 2\%. The Hubble constant inference using these effective bright sirens can be robust to astrophysical model error up to $\sim60\%$ change in galaxy mass function. We further observe that the number of included sources increases exponentially to achieve smaller statistical error. This again reflects the fact that deep redshift events offer poor constraining power due to the lack of supporting galaxy catalog information. In a more comprehensive analysis, the Bayesian inference framework is typically adopted, which considers the posterior of each GW event instead of using Gaussian error approximation. The analysis on this exponentially growing GW catalog would pose heavy computational strain with moderate payoff.

Finally, we comment on the overall magnitude of galaxy theoretical model error tolerance. The fiducial models explore effects from different error sources, and the realistic models highlight the transition between dark siren and effective bright siren. We discuss each aspect in turn. 

The fiducial models show that, in general, tolerance to $\beta$ is on the order of $\mathcal{O}(0.01)$ if the dark siren method were to achieve a Hubble measurement with a percent-level total error budget. This is expected to be applicable for realistic models as well, if localization errors are slightly larger than our assumed scenario, i.e., CE+LL+ET. Comparing \textit{\texttt{sparse}} and \textit{\texttt{realistic\_bbh\_rate}}, we see that increasing galaxy spacing or reducing density improves the rate model sharpness, which translates into more robustness against model bias by $\sim20\%$. While galaxy density is in reality fixed, more complex host galaxy model may achieve similar effect. Although in the \textit{\texttt{deepz}} simulation we also use high GW event occupation rate, the max $\beta$ tolerance is higher than the baseline \textit{\texttt{optimistic}} by $23\%$. This suggests that increasing high-precision galaxy catalog depth is very important for managing potential bias. This result thus highlights the challenge in controlling the bias in the GW dark siren Hubble measurement. Even for low-redshift universe ($z<0.1$), the Schechter function on the distribution of galaxy luminosity function or surface velocity dispersion function is only constrained with a parameter uncertainty on the order of $5\%$ with the SDSS dataset \cite{Choi}. Redshift evolution model uncertainties are also limited by the lack of deep redshift observations, and a percent-level knowledge can be challenging. 

We now discuss the implication from realistic model tests. Most notably, the error tolerance increases exponentially in the regime when the close-by effective bright sirens already provide sufficient statistical power to constrain the Hubble constant. For our particular condition, this shift occurs when the total error budget is between $1.4\%-2.0\%$. This result shows that, if GW localization produces sufficient effective bright sirens, the gain for including higher redshift faint sources is limited; while the statistical error is suppressed, susceptibility to bias quickly grows, and the total error budget does not improve. We note that this behavior is not present in fiducial model simulations. This is mostly because the fiducial model localization assumption is more conservative and we have yet to reach the effectively bright siren scenario. 

In our exploration using a clustered mock galaxy catalog, we observe that the error tolerance is better than the Poisson sampling scenario. This suggests that our majority model results will provide a conservative baseline to a full dark siren analysis using real observed galaxy catalogs.

\section{Conclusion and Future Work}
\label{chap:conclusion}
In this work, we use the Fisher information formalism to quantify potential bias in the GW dark siren measurement of the Hubble constant in the context of third-generation GW detector network. We use a mixture model between galaxy catalog and theoretical galaxy number density and simulate the full-sky Fisher information. For both galaxy survey and GW measurement parameters, we create variation models to explore the range of behavior. We assume a power law error in the galaxy mass function and find the maximum error tolerance given a range of total error budgets.

We found that the galaxy redshift error, galaxy completeness fraction (i.e. magnitude limit) and the GW angular localization error are significant factors that contribute to model error tolerance. In the case where GW dark sirens are not effectively localized and close-by sources do not provide the desirable Hubble measurement precision, the error tolerance trend can demonstrate sweet spots, beyond which including further and fainter GW events in the inference starts to impose harsher requirement on our knowledge of the galaxy mass function. Over the range of considered parameters, we notice that to achieve a total Hubble measurement budget of $1\%$, the maximum $\beta$ is on the order of $\mathcal{O}(0.01)$. In contrast, when well-localized sources (i.e. small measurement uncertainty) can already satisfy the target error budget, it is preferred to only use these effective bright sirens; this dramatically improves tolerance to astrophysical model uncertainties $\beta_\mathrm{max}\sim\mathcal{O}(1)$. This advantage is further strengthened by our additional test using realistic clustered galaxy catalog.

Our study thus shows the importance of jointly considering galaxy survey and GW measurement, and the simulation results can contribute to the decision for GW source selection strategy for dark siren inference of cosmological parameters.

We now discuss several directions for further investigation on this project. Firstly, we can improve on the galaxy catalog model with non-uniform galaxy hosting rates for GW depending on galaxy type (e.g. Ref. \cite{cao_hostgalaxy}). Different from an overall redshift evolution in GW occupation rate, this creates more local rates difference, and the smoothing from overlaying rate volume may be mitigated. This modification, similar to galaxy clustering, increases the ``spikiness'' of the Fisher information, and therefore may be expected to improve the dark siren robustness.

% Firstly, we can improve on the galaxy catalog model with galaxy clustering. On one hand, galaxies within the same group are closer together, and stronger overlaying of rate volume from each galaxy could lead to a more featureless rates model. On the other hand, the large scale structure ``troughs'' can be wider than the flat galaxy distribution separation, thus adding more large scale features. Which factor becomes dominate should be clarified with an improved model. 

Secondly, integration over different GW mass pairs and source location can be done explicitly. In this current approach, we select a random mass pair and treat it as an average case. Overall, the expected effect is a slight addition to the Fisher information, since the detector-frame mass distribution also changes with Hubble constant. However, if we assume that GW sources of all masses (within the detectable range of ground-based detectors) are evenly distributed in galaxies, the rates profile for each GW species can be quite similar. In our simulation, we have shown that sharp features provides the most information, and it may not be expected that smooth mass profile variation changes our estimate significantly. 

Moreover, the simulation results can be refined by considering larger simulation volume with high resolution. We note that the completeness fraction for more relaxed selection thresholds extends well beyond $z=2$. This creates artifacts in, e.g. the right edge of Fisher information in Figure \ref{fig_fid_fisher_info} and Figure \ref{fig_consis_IH} and potentially brings inaccuracies in the $\beta$ tolerance results. In our searched parameter space, we avoid threshold values that are too relaxed and include major contribution from higher redshift. In addition, the Fisher information typically drops well before this limit due to the smoothing of the rates model (see, e.g., Figure \ref{fig_fid_fisher_info}), indicating that our current calculation still captures the major effect. These regions are relatively far away from the most interesting parameter space of this work, and should not alter the overall trend and the main conclusions. 

\acknowledgements

Our research was supported by the Brinson Foundation, the Simons Foundation (Award Number 568762) and NSF Grants PHY-2309231 and PHY-2309211. 
The computations presented here were conducted in the Resnick High Performance Computing Center, a facility supported by Resnick Sustainability Institute at the California Institute of Technology.

This work has made use of CosmoHub.

CosmoHub has been developed by the Port d'Informació Científica (PIC), maintained through a collaboration of the Institut de Física d'Altes Energies (IFAE) and the Centro de Investigaciones Energéticas, Medioambientales y Tecnológicas (CIEMAT) and the Institute of Space Sciences (CSIC \& IEEC).
CosmoHub was partially funded by the "Plan Estatal de Investigación Científica y Técnica y de Innovación" program of the Spanish government, has been supported by the call for grants for Scientific and Technical Equipment 2021 of the State Program for Knowledge Generation and Scientific and Technological Strengthening of the R+D+i System, financed by MCIN/AEI/ 10.13039/501100011033 and the EU NextGeneration/PRTR (Hadoop Cluster for the comprehensive management of massive scientific data, reference EQC2021-007479-P) and by MICIIN with funding from European Union NextGenerationEU(PRTR-C17.I1) and by Generalitat de Catalunya.

\bibliographystyle{apsrev4-2}
\bibliography{ref}

\end{document}